\documentclass[reprint,amsmath,amssymb,aps,rmp]{revtex4-2}

\usepackage[utf8]{inputenc}
\usepackage{graphicx}
\usepackage{hyperref}
\usepackage[version=4]{mhchem}
\usepackage{bm}
\usepackage{upgreek}
\usepackage{physics}

\begin{document}

\title{Long-range electrostatics in atomistic machine learning: a physical perspective}

\author{Federico Grasselli}
\affiliation{Dipartimento di Scienze Fisiche, Informatiche e Matematiche, Universit\`a degli Studi di Modena e Reggio Emilia, 41125 Modena, Italy}
\affiliation{CNR-NANO S3, 41125 Modena, Italy}

\author{Kevin Rossi}
\affiliation{Department of Materials Science and Engineering, Delft University of Technology, 2628 CD Delft, The Netherlands}
\affiliation{Climate Safety and Security Centre, TU Delft The Hague Campus, Delft University of Technology, 2594 AC The Hague, The Netherlands}

\author{Stefano de Gironcoli}
\affiliation{Scuola Internazionale Superiore di Studi Avanzati (SISSA), 34136 Trieste, Italy}

\author{Andrea Grisafi}
\email{andrea.grisafi@sorbonne-universite.fr}
\affiliation{Physicochimie des Électrolytes et Nanosystèmes Interfaciaux, Sorbonne Université, CNRS, F-75005 Paris, France}

\begin{abstract}
The inclusion of long-range electrostatics in atomistic machine learning (ML) is receiving increasing attention for achieving quantum-mechanical accuracy in predicting a wide range of molecular and material properties. However, there is still no general prescription for how long-range physical effects should be incorporated into a model while preserving the well-established locality principles underlying most transferable ML representations.

Here, we provide a physical perspective on this problem by discussing how distinct contributions to a system's electrostatics can be captured through different learning paradigms. Specifically, we distinguish between local charge models, which rely either on explicit charge-density decompositions or implicit auxiliary variables, and models in which a notion of nonlocality is deliberately introduced, either via self-consistent procedures or by using nonlocal descriptors and learning architectures. We further address the related aspect of incorporating finite-field effects through coupling to the system's polarization, relevant for applications involving an external electric bias. We conclude by discussing implications for simulations of electrochemical interfaces, where long-range electrostatics are essential to capture the interplay between charge redistribution, interfacial dynamics, and ionic screening, as well as for ionic transport phenomena, which, although less explored, appear far less sensitive to their inclusion.
\end{abstract}

\maketitle

\tableofcontents

\section{Introduction}

Machine-learning (ML) methods are rapidly revolutionizing the field of  atomistic simulations by making it possible to attain the accuracy of first principles approaches while reducing the computational burden  by 3-5 orders of magnitude. The vast majority of these methods rely on the principle of electronic nearsightedness~\cite{prod-kohn05pnas}, implying that a local decomposition of the physical target is adopted to ensure the transferability of the model across diverse chemical and structural patterns. In practice, this locality assumption is enforced in the ML model by adopting input representations of the atomic coordinates defined via spherical windows of a given real-space cutoff around the atoms of the system~\cite{musil_chem_rev}. 
As a result, local ML models naturally satisfy the extensive character of the electronic energy, and their evaluation scales linearly with the number of atoms, in sharp contrast to routine \textit{ab initio} methods, whose cost grows superlinearly due to the explicit treatment of the electronic structure.

The assumption of locality has proven to be generally valid in the simulation of bulk materials, where the effect of long-range interactions is averaged out when converging most of the thermodynamic properties of the system~\cite{Kocer2022}. 
Conventional static distributions, such as the radial distribution function, are usually reproduced with high accuracy by short-range ML potentials up to interatomic distances of several~\AA.  
A major exception arises, however, when considering the long-wavelength limit (i.e., vanishing wavevector, $k\!\to\!0$) of \emph{electric} response properties.  
For instance, the static charge–charge structure factor, defined from the Fourier transform of the charge density $\rho_Q$ as
\begin{equation}
    S_{QQ}(\boldsymbol{k}) \propto \langle \tilde{\rho}_Q(\boldsymbol{k})\,\tilde{\rho}_Q(-\boldsymbol{k}) \rangle,
\end{equation}
vanishes as $k^2$ in charged-particle fluids with truly long-range Coulomb interactions, but remains finite in short-range models.  
Although strict charge neutrality ensures that $S_{QQ}(0)=0$ for any finite simulation box, even for short-range models, the crucial difference emerges in the \emph{limit} $k \!\to\! 0$: long-wavelength charge-density fluctuations are suppressed by perfect screening only in systems with long-range Coulomb interactions~\cite{copestake1982charge}.
Analogously, only models that include explicit long-range interactions correctly reproduce the $k \!\to\! 0$ behavior of the longitudinal component of the dipole--dipole correlation function, determining the macroscopic dielectric response of the system, as exemplified in the case of bulk liquid water~\cite{Gao2022}.

While these exceptions can be seen as narrowly defined targets, the lack of long-range physics in ML models becomes particularly problematic in the description of atomistic interfaces or finite clusters/molecular systems. In this case, the geometrically unbalanced (i.e., anisotropic) distribution of the electrostatic fields produces a macroscopic  effect in the equilibrium properties of the system that must be properly account for, as shown in the context of liquid-vapor interfaces~\cite{Niblett2021}. This aspect is tremendously magnified when considering the simulation of electrified interfaces, as well as systems including generic electronically conductive materials, where electronic charge transfers and long-range electronic polarizations also come into play~\cite{Sundararaman2022,gross2023}.

\begin{figure}
    \centering  \includegraphics[width=1\linewidth]{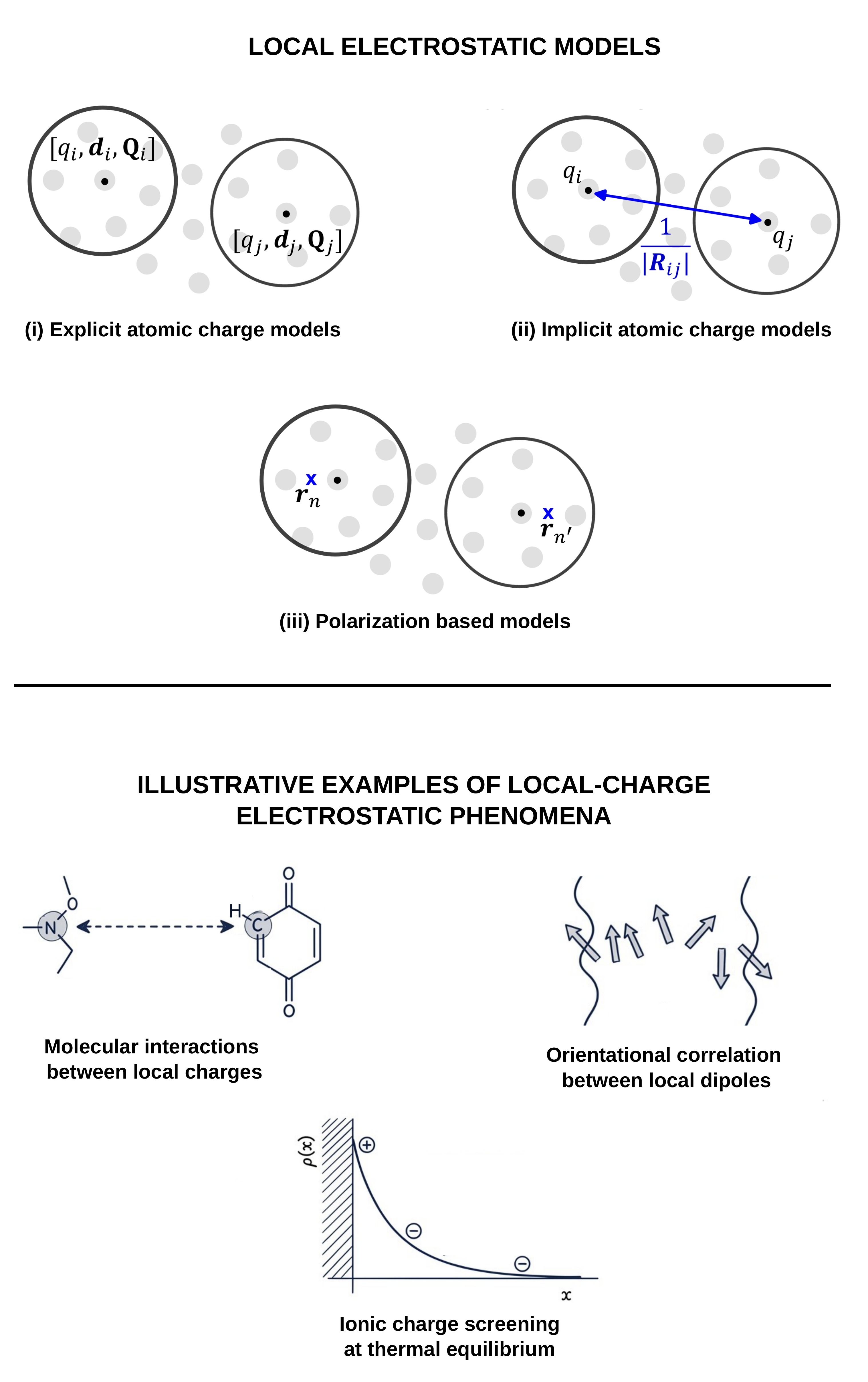}
    \caption{\textit{Top}: families of local electrostatic models in atomistic machine learning: (i) quantum-mechanical moments derived from an atomic partitioning of the charge density are learned explicitly, (ii) atomic charges are treated as auxiliary variables and implicitly inferred upon learning electronic energies and/or global dipoles, (iii)  {a consistent} representation of the polarization vector in periodic systems is adopted through learning of Wannier centers, or atomic dipoles treated as auxiliary variables. \textit{Bottom}: {Illustrative examples of local electrostatic phenomena: the interaction between local charge distributions in distant molecules; the orientational (thermodynamic) correlations between local dipoles; the ionic charge screening at thermal equilibrium}.}
    \label{fig:method-1}
\end{figure}

In this perspective, we outline main research directions towards the inclusion of electrostatic interactions into ML models. In so doing, we will not touch on any performance metrics or specific technical contraption of neural-network architectures, but rather focus our discussion on the core physics and fundamental principles adopted by different machine-learning methodologies. Specifically, we will differentiate the problem of learning the long-range physics in the system via local representations of charge density, thus accounting for permanent electrostatics and local, i.e., within-cutoff, polarization phenomena, where long-range polarization can be captured at most as a mean, dataset-dependent statistical effect, from that of incorporating a notion of nonlocality into the model, necessary to recover long-range, structure-dependent polarizations and electronic charge transfers. 
We will further discuss the related problem of dealing with the application of a finite electric field, often included to simulate an external electric bias through the coupling with the system's polarization.
We will conclude showcasing example applications concerning the role of long-range physics in the simulation of electrochemical interfaces, as well as in the calculation of ionic transport properties. Atomic units will be used all throughout.

\section{Electrostatics from local charge models}\label{sec:local-models}

Over the years, the modeling of electrostatic interactions via machine-learning approaches has seen a number of substantially different developments~\cite{Anstine2023}. This heterogeneity ultimately derives from seeking a compromise between the inherent locality of most transferable ML methods and the long-range nature of the electrostatic energy, $U_\text{ele}$. At a general level, $U_\text{ele}$ is expressed as the self-interaction of the charge density $\rho_Q$ via the (nonlocal) Coulomb potential:
\begin{equation}\label{eq:electro}
    U_\text{ele} = \frac{1}{2} \int \mathrm{d}\boldsymbol{r}\, \int \mathrm{d}\boldsymbol{r}' \, \frac{\rho_Q(\boldsymbol{r})\rho_Q(\boldsymbol{r}')}{|\boldsymbol{r}-\boldsymbol{r}'|}\, ,
\end{equation}
where $\rho_Q$ is defined from the sum of nuclear charges $\{Z_i\}$ and the electron density $\rho_e$, i.e., $\rho_Q(\boldsymbol{r}) = \sum_i Z_i\delta(\boldsymbol{r}-\boldsymbol{R}_i) + \rho_e(\boldsymbol{r})$, with $\boldsymbol{R}_i$ the atomic positions.

A practical approach to circumvent the nonlocality of $U_\text{ele}$ consists in relying on a discrete decomposition of $\rho_Q$ that is, more or less indirectly, inferred by local ML models. 
In particular, given the local environment representation $X_i$ of an atom $i$ -- defined from the set of atomic coordinates $\{\boldsymbol{R}_j\}$ falling within a given spherical cutoff $r_\text{cut}$ -- any local component $c$ of the charge density can be generally predicted as $c(X_i) = f_{\boldsymbol{\theta}}(X_i)$, with $\boldsymbol{\theta}$ denoting the ML fitting parameters  and $f$ a possibly nonlinear function expressing the learned data-driven map between $c$ and the input coordinates. As we will see, the components $c$ can be identified as partial atomic charges, atomic multipoles, Wannier-center charges, or generic electron-density coefficients on a given atomic basis.
In the absence of a self-consistent update of $\rho_Q$, any local machine-learning methodology that follows this rationale inherently neglects polarization effects induced by far-field structural rearrangements,
implying that 
\begin{equation}\label{eq:locality-constraint}
    \frac{\partial c(X_i)}{\partial \boldsymbol{R}_k}=0\, ,\quad \text{for}\ |\boldsymbol{R}_k-\boldsymbol{R}_i| > r_\text{cut}\, .
\end{equation}
Despite this limitation, local charge models are, in principle, capable of predicting any charge variation determined by the structural changes occurring
within the (possibly wide) spatial range covered by the atomic environment cutoff. In most cases, this is sufficient to provide the leading contribution to the Coulomb interactions in the system, affecting the configurational and binding energy in molecular frameworks, as well as key thermal equilibrium properties of  material interfaces. Moreover, it should be noted that as the model can be trained not only on the quantum-mechanical charges directly, but also on more or less nonlocal physical quantities, e.g., electronic dipoles and energies, the data-driven determination of the charge-density components can effectively incorporate a statistical notion of long-range polarization, providing that this mirrors the type of long-range physics encountered during the training procedure.

In this section, we discuss how  different local-charge models can be built from various choices of quantum-mechanical learning targets. As summarized in Fig.~\ref{fig:method-1}, these include: (i) explicit electronic-charge partitionings, {where the learning targets are identified as quantum-mechanical atomic charges and multipoles}, (ii)~implicit charge partitionigs, {where the atomic charges act as auxiliary variables and are implicitly} inferred via the learning of {quantum-mechanical global} multipoles, {e.g., molecular dipoles,}  and/or electronic energies, (iii) Wannier centers {and atomic dipoles} {representations} accounting for  {a consistent} definition of the polarization vector in periodic systems. For the sake of simplifying notation and making the discussion uniform when comparing different methods, we will report single-sample loss functions using an absolute norm metric, even though many models actually adopt a square-norm choice.

\subsection{Explicit charge models}

The most straightforward strategy to build a charge-aware ML model consists in directly learning an atomic decomposition of the quantum-mechanical $\rho_Q$ over a set of atomic partial charges $\{q_i\}$ -- see Fig.~\ref{fig:method-1}-(i). These decompositions derive from a long-standing tradition in quantum-chemical analysis~\cite{bade02book}, following the need of assigning a unique charge value to each atom in the system.  Early ML models that incorporated this approach into machine-learning potentials have been presented in Refs.~\cite{Artrith2011,Morawietz2012}, opening the so-called third generation of high-dimensional neural-network potentials (3D-HDNNP). 
In this case, the electronic energy is expressed via a formal separation between short-range and electrostatic energy. Specifically, the atomic charges $q_i$ are first computed from a Hirshfeld partitioning of the DFT charge density~\cite{hirshfeld_bonded-atom_1977}. These charges are then used to provide a point-charge approximation to the electrostatic energy of Eq.~\eqref{eq:electro} and, in turn, to define the short-range contribution to the total electronic energy $U$ as $U_\text{SR} = U - U_\text{ele}$.   Separate NN models based on symmetry-functions representations of the local atomic environments~\cite{behl-parr07prl} are independently trained on the computed charges and $U_\text{SR}$ values, so that the two contributions to the total atomic forces are finally added together upon prediction. When considering the evaluation of electrostatic forces, in particular, the dependence of the $q_i$ upon the atomic coordinates -- acquired through the environment representation used as input -- is also taken into account to enforce conservation of the total energy. Applications of 3D-HDNNP have been shown to model interactions in large nanostructured zinc-oxide systems ~\cite{Artrith2011} and water dimers~\cite{Morawietz2012}, where explicit atomic charges can effectively be predicted by a local model. 
Beyond ML potentials, we note how learning models of atomic partial charges have also been proposed in the context of including an environment-dependent prediction of $q_i$ within classical force-fields, where the notion of point-charges is traditionally built-in~\cite{Bleiziffer2018}. The domain of applications of this approach concerned the prediction of organic molecular properties -- namely heats of vaporization and hydration free energies -- demonstrating how the prediction of atomic charges can be generalized over a set of chemically diverse compounds.

One of the limitations of atomic charge models is that they neglect the local anisotropy of the underlying physical charge density, thereby assuming that the effect of dipolar and higher-order atomic multipoles, $\mathbf{M}_i$, can effectively be absorbed into the short-range contribution to the electronic energy. In this regard, models that express the electrostatic energy via ML approximations of $\mathbf{M}_i$ up to second order, i.e., atomic charges ($q_i$), dipoles ($\boldsymbol{d}_i$), and quadrupole moments ($\mathbf{Q}_i$), were introduced since the early days of atomistic machine learning  with case-studies reported for water~\cite{Handley2009} and ethanol~\cite{Mills2011}. 
A  transferable evolution of this idea was developed in Ref.~\cite{Bereau2015}, where the quantum-mechanical atomic multipoles are extracted from Gaussian distributed multipole analysis~\cite{stone+97book,Stone1985} (GDMA), and used to feed a kernel-based ML model that learns the difference between $\mathbf{M}_i$ and inexpensive baseline values of atomic multipoles computed through a Voronoi partitioning of pretabulated atomic densities. 
The approach has been later used to include ML-based predictions of $\mathbf{M}_i$ within a physics-based  force-field parametrization  of the molecular interatomic potential, achieving chemical accuracy across a diverse dataset of gas-phase dimers{, including DNA-base and amino-acid pairs}~\cite{Bereau2018}.

Equivariant graph neural network (GNN) improvements of this method that account for the transformation properties of $\mathbf{M}_i$ under rotations have been presented in Ref.~\cite{Thuerlemann2022}, leveraging the multipolar decomposition of $\rho_Q$ to reproduce accurate three-dimensional maps of electrostatic potential profiles. This investigation concerned model training, validation, and testing across hundreds of thousands of molecules with up to 20 heavy atoms (H, C, N, O, F, S, Cl).

While these types of atomic decompositions provide a straightforward strategy to include information about the charge distribution of the system, they fundamentally rely on an arbitrary partitioning choice. A different possibility is that of directly learning the electron density $\rho_e$ as computed from density-functional theory (DFT) and other electronic-structure methods. In fact, a number of ML developments have been presented over the last decade that are specifically designed to provide a representation of $\rho_e$ as a three-dimensional scalar field. In this case, two main directions have been independently followed that either expand the scalar field over a set of basis functions~\cite{broc+17nc,gris+19acscs, Rackers2023}, or directly learn $\rho_e$ on a real-space grid~\cite{Focassio2023, Jorgensen2022, Koker2024, Feng2025}.
Because of the much higher complexity of the learning target, however, these approaches have not, so far, been extensively explored to compute  electrostatic interactions.  

A paradigmatic example of electron-density representation -- which can be seen as bridging the gap between multipolar expansions and the full $\rho_e$ -- consists in approximating the electron density on a linear atom-centered basis:
\begin{equation}\label{eq:RI}
    \rho_e(\boldsymbol{r}) \simeq\sum_{i=1}^N \sum_{\nu} c^i_{\nu}\,  \phi_\nu(\boldsymbol{r}-\boldsymbol{R}_i)\, ,
\end{equation}
where $\phi_\nu$ are the atomic basis functions, and $c^i_\nu$ are the expansion coefficients that can be predicted by local ML models~\cite{gris+19acscs}.
Expansions as in Eq.~\eqref{eq:RI} are commonly derived from resolution of the identity (RI) approximations in quantum chemistry~\cite{ren+12njp,Stoychev2017}, and are therefore available in many electronic-structure programs. When it comes to the ML prediction, each atom in the system is, in this case, associated with a number of electron-density coefficients  of the order of~$ 10^2$, rather then a small number of atomic multipoles, adding a substantial linear prefactor to the computational cost at inference in exchange for the acquired expressiveness. 
An even more pronounced increase of dimensionality is found when evaluating $\rho_e$ on real-space grids, although no approximation is introduced in the reference electron-density representation in this case.

By and large, it is debatable whether it is worth adopting ML methods of the  electron density to represent long-range interactions. 
Although atomic multipoles can be explicitly introduced in the charge density approximation, a distinction should be made between these and the \textit{global} physical multipoles, e.g., molecular and interface dipoles, which directly govern the far-field electrostatics. 
While a separate modeling of $U_\text{SR}$ can effectively absorb any missing contribution to the Hartree energy, this does not automatically guarantee that the global  electrostatics in the system is accurately reproduced. 
In this context, ML models capable of accessing the full electrostatic energy at once through electron-density predictions could promise a more robust and reliable solution.

\subsection{Implicit charge models\label{sec:implicit}}

A practical approach to compensate the inherent arbitrariness of explicit atomic-charge partitioning schemes is that of interpreting the set of $\{q_i\}$ as auxiliary variables that can be indirectly inferred through the learning of well-defined {global} physical quantities, as sketched in Fig.~\ref{fig:method-1}-(ii). One of the first ML methods -- TensorMol~\cite{yao+18cs} -- that applied this idea to molecular systems proposed to train the model on the quantum-mechanical dipole moment $\boldsymbol{d}$. In finite and charge-neutral systems, this can be simply fitted as $\boldsymbol{d} = \sum_{i=1}^N q_i\, \boldsymbol{R}_i$.
In this case, the learning model establishes the statistical correlation between the reference quantum dipoles, $\boldsymbol{d}^\text{ref}$, and the local ML representation of the partial atomic charges. This corresponds to minimizing the loss function:
\begin{equation}\label{eq:dipole}
\mathcal{L}({\boldsymbol{\theta}}) \sim \sum_{\alpha=1}^3\left|\sum_{i=1}^N q_i({\boldsymbol{\theta}})\, R_{i,\alpha}-d^\text{ref}_\alpha\right|\, ,
\end{equation}
where $\alpha$ are the Cartesian indexes. In practice, the position vectors $\boldsymbol{R}_i$ provide a global metric to combine the local atomic features during the training procedure, necessary to find the optimal partitioning of the charge density to reproduce the molecular dipole moment. Crucially, however, this global coupling does not introduce a nonlocal character at inference, where the atomic charges are predicted based solely on the local environment information. Similarly to Hirshfeld charge models, an approximation of the electrostatic energy is then computed from the fitted $q_i$, while the missing contribution to the electrostatic energy is assumed to be short ranged and effectively incorporated into $U_\text{SR}$. TensorMol was validated for water clusters, large organic molecules, and DNA-pair interactions.
Although not directly related to the prediction of long-range energy, we note how models that learn molecular dipoles by adopting a loss function as in Eq.~\eqref{eq:dipole} have shown to be suitable in representing zwitterionic molecular chains (MuML~\cite{veit+20jcp}), where a decomposition of $\boldsymbol{d}$ in terms of atomic charges can naturally capture, by the model construction, extended (permanent) charge separations over arbitrary large distances.

\begin{figure}[h!]
    \centering
    \includegraphics[width=1\linewidth]{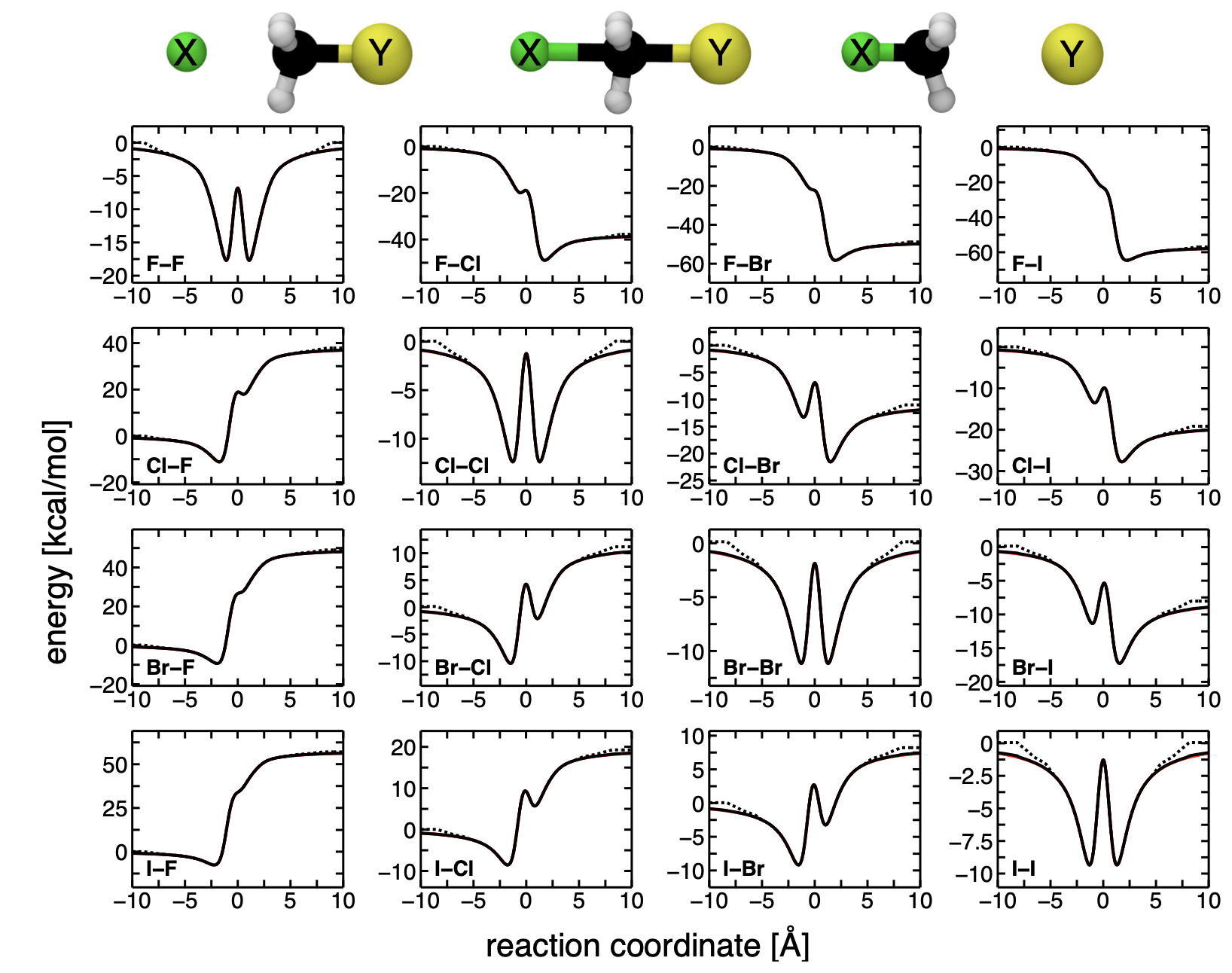}
\caption{{Minimum energy paths for various S$_\text{N}$2 reactions, calculated using the PhysNet model with (solid black
line) and without (dotted black line) explicit long-range interactions. Reproduced from Ref.~\cite{unke-meuw19jctc} with permission from the American Chemical Society. Copyright 2019 American Chemical Society.}}
    \label{fig:physnet}
\end{figure}

Following the rationale of inferring the atomic charges from computed electrostatic quantities, evolutions of the ideas previously reported, still applied to molecular systems, have been introduced in the PhysNet model~\cite{unke-meuw19jctc}. 
Here, a message-passing NN architecture is simultaneously trained on the electronic energy and total dipole model by minimizing a multi-targeted loss function that additionally enforces the conservation of the total charge~$Q$:
\begin{equation}\label{eq:multi-loss}\begin{split}
\mathcal{L}({\boldsymbol{\theta}}) &\sim w_U\left|U_\text{ML}(\{q_i({\boldsymbol{\theta}})\})-U_\text{ref}\right| \\&+ w_d\sum_{\alpha=1}^3\left|\sum_{i=1}^N q_i({\boldsymbol{\theta}})\, R_{i,\alpha}-d^\text{ref}_{\alpha}\right| \\&+ w_q\left|\sum_{i=1}^Nq_i({\boldsymbol{\theta}}) - Q\right|\,  + (\text{atomic force terms})
\end{split}\end{equation}
with $w$ weighting the relative contribution of the various terms. The total electronic energy is here defined as
\begin{equation}
    U_\text{ML}(\{q_i\}) \sim U_\text{SR} + \sum_{i\ne j} \frac{q_iq_j}{|\boldsymbol{R}_i-\boldsymbol{R}_j|}f_\text{damp}(R_{ij})\, ,
\end{equation}
where $f_\text{damp}(R_{ij})$ is a damping factor at short interatomic distances.
In so doing, PhysNet fits the atomic charges to simultaneously reproduce both the total dipoles $\boldsymbol{d}$ and the total electronic energies, hence bypassing the need to handle the short-range part of the electronic energy via a separate learning task. In this context, an additional global coupling between atomic environments is provided by the Coulomb potential entering the definition of electrostatic energy, which introduces a nonlocal physical constraint in the atomic-charge partitioning. We remark, however, that this nonlocality only serves to provide a statistical map between the electrostatic energy and the local structural features, and no far-field information is used to predict the atomic charges, as formalized in Eq.\eqref{eq:locality-constraint}. PhysNet provided state-of-the-art performance on the QM9, MD17 and ISO17 benchmark datasets~\cite{unke-meuw19jctc}. Moreover, the method was validated in the context of  S$_\text{N}$2 reactions of methyl halides with halide anions{, where the inclusion of long-range interactions has shown to be essential to reproduce the correct dissociated limits (Fig.~\ref{fig:physnet}),} and to prove its transferability across solvated protein fragments. 

Another development that adopts a multi-targeted loss function similar to Eq.~\eqref{eq:multi-loss} has been presented in the AIMNet2 model~\cite{Anstine2025}. In addition to total energies and molecular dipoles, the model is also trained on molecular quadrupole moments, further enforcing the physical constraints of the fitted atomic charges. Its applicability was demonstrated from simple organics molecules to larges ones, including so defined “exotic” element-organic bonding.
Moreover, the atomic charges become an integral part of the descriptor, and are iteratively updated during the message-passing procedure to conserve the total molecular charge {$Q$, thus allowing for treating open-shell and ionic systems. Specifically, the atomic charges at step $n$ of iteration are given~by:
\begin{equation}\label{eq:fukui}
    q_i = \tilde{q}_i + \frac{f_i}{\sum_{j=1}^N f_j} \left(Q - \sum_{j=1}^N \tilde{q}_j\right) 
\end{equation}
where $\tilde{q}_i$ are the atomic charges at step $n-1$, while $f_i$ represent some learnable parameters mimicking the atomic Fukui functions~\cite{Zubatyuk2021}. We note that while this procedure introduces a global charge update at inference, this only serves to enforce conservation of the total charge rather than to explicitly account for the effect of far-field structural changes in the system, so that it can still be considered as a ($Q$-conserving) local-charge model for the sake of the present discussion. Overall, AIMNet2 provides a scalable machine-learning framework for fast quantum-chemistry–level molecular simulations.} Notably, while training was limited on gas-phase molecules, the learned information could be extrapolated in the condensed phase to perform the molecular dynamics simulation of weakly interacting molecular liquids. However, this level of transferability is likely to be lost when considering arbitrary condensed-phase systems, where the many-body character of interactions breaks the assumed molecular additivity of the total electronic energy.

\begin{figure}[h!]
    \centering
    \includegraphics[width=0.9\linewidth]{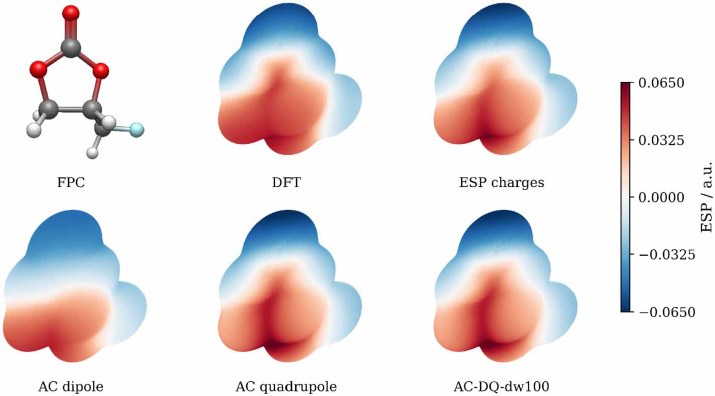}
    \caption{{Electrostatic potential of fluoropropylene carbonate calculated on the 0.001 a.u. density isosurface. The bottom panel shows the potential from atomic charges predicted with PiNet2-based dipole and quadrupole models. Reproduced from Ref.~\cite{Muuga2026}, licensed under CC BY~4.0.}}
    \label{fig:eps_map}
\end{figure}

{In addition to the atomic charges, we mention how the atomic dipoles $\boldsymbol{d}_i$ have also been suggested as possible learnable auxiliary variables within an equivariant GNN, i.e., PiNet2~\cite{Li2025PINN}, used to reproduce the global molecular dipole using the formal oxidation-state (OS) atomic charges ($q^\text{OS}_i$) as a baseline:
\begin{equation}
    \boldsymbol{d} = \sum_{i=1}^N q_i^\text{OS} \boldsymbol{R}_i + \boldsymbol{d}_i(\boldsymbol{\theta})\, .
\end{equation}
}
{More recently, a study conducted on the QM9 and SPICE datasets derived from an evolution of PiNet2 has shown the central importance of molecular quadrupole moments to obtain an atomic-charge partitioning capable of reproducing chemically-accurate molecular electrostatic potential (MEP) profiles~\cite{Muuga2026}. Notably, it was shown how an implicit charge model trained only on quadrupole moments produces more accurate MEP predictions that models trained exclusively on molecular dipoles (Fig.~\ref{fig:eps_map}), while an optimal level of accuracy can be achieved by including both $\boldsymbol{d}$ and $\boldsymbol{Q}$ in the training data.}

\begin{figure}[h!]
    \centering
    \includegraphics[width=0.8\linewidth]{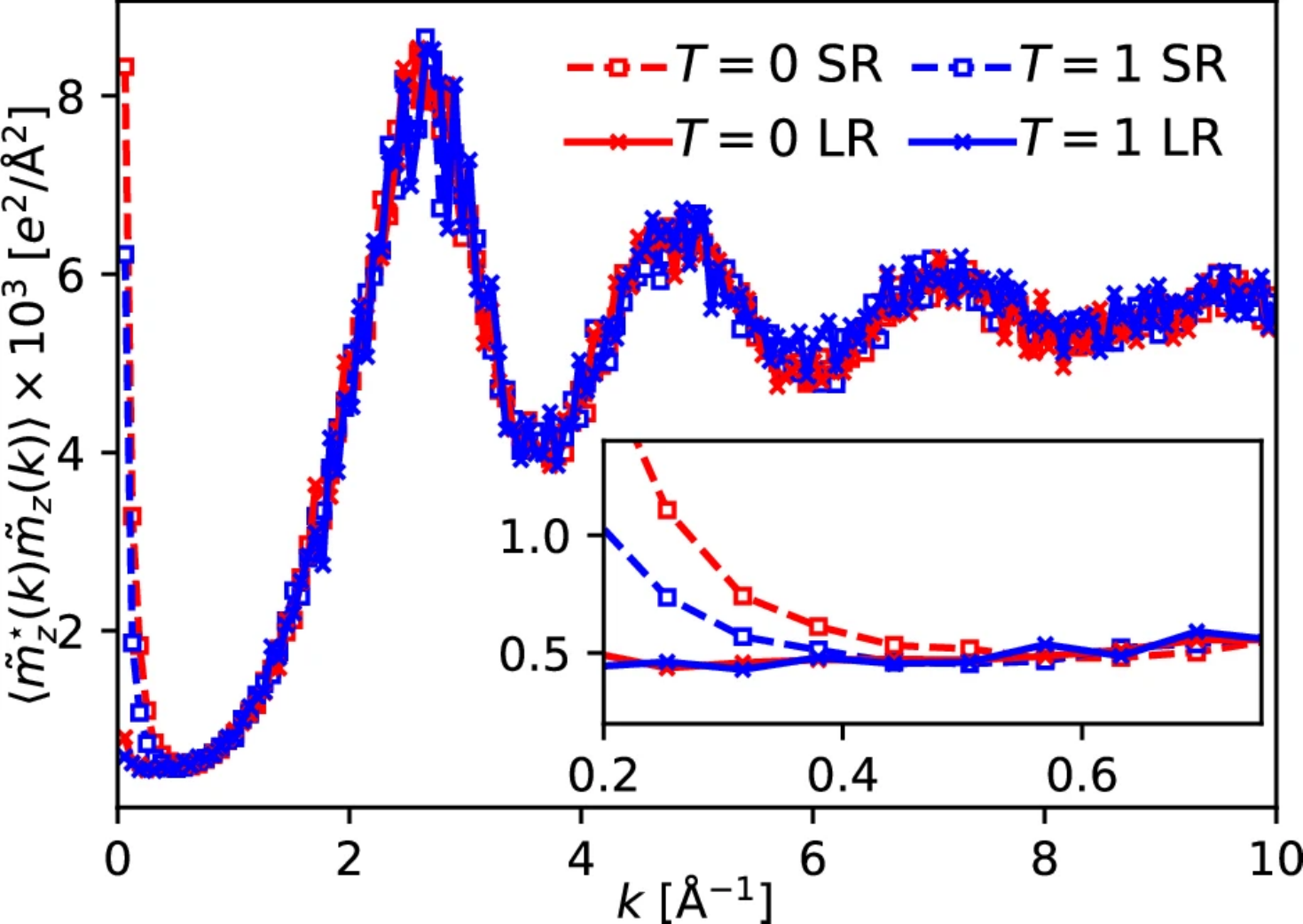}
    \caption{{Predicted dipole density correlations of bulk liquid water in reciprocal space, using short-range (SR) or long-range (LR) models and with or without massage passing layers ($T = 0$ or $T = 1$). The inset shows a zoom-in of the small--$k$ behavior. Reproduced from Ref.~\cite{Cheng2025}, licensed under CC BY 4.0.}}
    \label{fig:corr_water}
\end{figure}

Beyond gas-phase molecules, natural extensions of implicit local-charge models that can also deal with condensed phases are realized by incorporating the periodicity of the simulation cell in the approximation of the electrostatic energy. A notable example is the latent Ewald summation (LES) model~\cite{Cheng2025,King2025,kim2025universal}. Following standard periodic formulations of the Coulomb potential, the electrostatic contribution to the total electronic energy is here expressed through the reciprocal space formula:
\begin{equation}\label{eq:ewald-rec}
    U(\{q_i\}) = U_\text{SR} + \frac{2\pi}{\Omega} \sum_k \frac{e^{-\frac{1}{2}\left(\sigma k\right)^2}}{k^2}\sum_{ij} q_iq_j\,  e^{-i\boldsymbol{k}\cdot(\boldsymbol{R}_i-\boldsymbol{R}_j)}\, ,
\end{equation}
with $\sigma$ the Gaussian width used to provide a smooth representation of the atomic charge densities and $\Omega$ the cell volume. Here, local environment features are once again adopted for the representation of the atomic charges $q_i$, which, in turn, are used as (latent) auxiliary variables of a message-passing NN architecture. From Eq.~\eqref{eq:ewald-rec}, a loss function that contains computed quantum data for electronic energies and atomic forces is then minimized to find the optimal set of $\{q_i\}$ underlying the periodic charge partitioning. In so doing, LES can learn electrostatic interactions in a similar spirit to what achieved by PhysNet in the case of finite molecular systems. 
In practice, the model has shown the capability of reproducing binding energies between charged and polar molecules embedded in periodic boxes, {the small--$k$ behavior of the dipole-dipole correlation in liquid water (Fig.~\ref{fig:corr_water}),} the orientational screening of surface dipoles formed at water/vacuum interfaces~\cite{Cheng2025}, as well as ionic screening effects at  solid/electrolyte interfaces~\cite{King2025} -- all examples requiring a treatment of local charge distributions. 

Interestingly, however, LES has also been used in the context of predicting global charge-transfer effects in datasets that are deliberately designed to present nonlocal charge rearrangements~\cite{Ko2021,King2025}. This may appear as surprising given that the model does not possess a long-range charge update at inference that would overcome the locality criterion of Eq.~\eqref{eq:locality-constraint}. Arguably, however, this possibility can be attributed to the limited size and conformational variability contained in each the individual datasets, exemplifying how local-charge models can effectively incorporate a statistical notion of nonlocality in the predicted atomic charges as long as the long-range structural variations included in the training set correlate with the distribution of reference electronic energies.

{A similar implicit atomic-charge model compatible with periodic boundary conditions, which additionally deals with dispersion interactions and adopts a different treatment of the real-space contribution in the Ewald summation method, has also recently been developed~\cite{zaverkin2025transferable}.}

As a final remark, we note that in all models previously summarized 
message passing can effectively enhance the description of intermediate-range correlations, yet it does not introduce genuine nonlocality in the charge distribution. 
Message passing in graph neural network potentials is often described as a way to propagate information beyond the local atomic neighborhood~\cite{Batatia2022}, but it remains fundamentally a short-range mechanism: each message-passing layer transmits information only within a finite cutoff $r_\text{cut}$, and stacking multiple layers effectively increases the cutoff to $r^\text{eff}_\text{cut} > r_\text{cut}$. Nevertheless, the increase is constrained by the connectivity of the cutoff graph: for instance, the model cannot propagate information between regions without intermediate atoms, as numerically verified for vacuum-separated biodimers in Ref.~\cite{rumiantsev2025learning}. 
Since the atomic features exchanged through message passing are computed from fixed, local descriptors, their values do not depend consistently on the global configuration of the system. 
Consequently, such architectures cannot reproduce true polarization effects beyond $r^\text{eff}_\text{cut}$, where the electronic response of one region may adapt to the instantaneous charge distribution of distant regions. In practice, the effective message-passing cutoff can extend up to a few nanometers, which in many cases is enough to encompass most of the long-range physics in the system~\cite{Anstine2025}, although increasing the number of message-passing layers beyond roughly six is found to become computationally prohibitive~\cite{fuchs2025learning}.
However, this extension is mostly formal and highly conditional. As highlighted in Ref.~\cite{nigam2022unified}, when using a two-body-to-two-body message passing at a fixed $r_{\text{cut}}$ -- which would naively suggest an effective cutoff $r^\text{eff}_\text{cut} = 3 r_\text{cut}$ -- the error obtained on the electrostatic energy of a randomized NaCl structure with a two-body reference potential is far larger than the error of a truly $3\, r_\text{cut}$ model built with two-body representations. Thus, message passing does not meaningfully replace an explicit increase of the cutoff when aiming to capture longer-range physics.

\subsection{ {Polarization-based models}\label{sec:wannier}}
A substantially different paradigm with respect to what previously discussed can be encountered when relying on the macroscopic polarization of a periodic system to determine the local charge partitioning -- see Fig.~\ref{fig:method-1}-(iii).
An infinitely repeated system is not simply a ``big molecule'': its topology is fundamentally different, and with it, the very definition of macroscopic polarization must change. This conceptual shift was first recognized in 1992 by Resta~\cite{resta1992theory}, who showed that only the \textit{differences} of macroscopic polarization could be measured, not an absolute polarization. Soon after,  the problem was reformulated by King-Smith and Vanderbilt using Berry-phase topological arguments~\cite{king1993theory}, laying the foundation of what is now known as the \textit{modern theory of polarization} (MTP). 
In this framework, two key features emerge: (\textit{i}) polarization $\boldsymbol{P}$ in a periodic system forms a lattice, and is thus defined only modulo a ``quantum of polarization'', i.e., the addition to $\boldsymbol{P}$ of $e \boldsymbol{T}/\Omega$, where $e$ is the charge quantum, $\Omega$ is the unit-cell volume, and $\boldsymbol{T}$ is \textit{any} lattice vector, does not change any physically observable quantity. (\textit{ii}) $\boldsymbol{P}$ cannot be written as the first moment of the total charge density $\rho_Q(\boldsymbol{r})$, due to the ill definition of the position operator in periodic systems, 
\begin{equation}
    \boldsymbol{P} \neq \frac{1}{\Omega} \int_\Omega \mathrm{d}\boldsymbol{r}\, \boldsymbol{r} \, \rho_Q(\boldsymbol{r})\, . %
    \label{eq:pol_neq_rxrho}
\end{equation}
The RHS side of Eq.~\eqref{eq:pol_neq_rxrho} can be used to compute the macroscopic polarization of the extended system only when the system consists of non-overlapping, independent moieties carrying well-localized charges or dipoles (i.e., the so-called Clausius-Mossotti limit). Unfortunately, this is almost never the case, especially in insulating systems characterized by covalent bonds \cite{resta2007theory}. Point (\textit{ii}) is distinct from the indeterminacy associated with the polarization quantum, which arises even in systems composed solely of classical point charges. 

A discrete charge model compliant with MTP relies on the fact that the electronic part of the macroscopic polarization of an insulating system can be expressed (modulo a quantum of polarization) in terms of the positions of the Wannier centers (WCs), $\boldsymbol{r}_n = \langle w_n | \hat{\boldsymbol{r}} | w_n \rangle$, which represent the expectation values of the position operator computed over the Wannier functions $w_n(\boldsymbol{r}) = \langle \boldsymbol{r} | w_n\rangle$:
\begin{equation}\label{eq:wannier-pol}
    \boldsymbol{P}_{el} =  -\frac{1}{\Omega} \sum_n \boldsymbol{r}_n \, .
\end{equation}
To obtain the total polarization one must add the contribution from the atomic (pseudo)cores, $\boldsymbol{P} = \boldsymbol{P}_{el} + \frac{1}{\Omega} \sum_{i=1}^{N} Z_i \boldsymbol{R}_i $, where $Z_i$ is the core integer charge. The number of Wannier functions equals the number of occupied bands, since the Wannier representation is a unitary transformation on the manifold of occupied Bloch states, similarly to orbital localization strategies commonly applied in quantum chemistry~\cite{lowd1956pr}.  
Each WC can be associated with a localized electronic orbital, thus allowing one to assign a local dipole contribution to specific atoms, bonds, or structural motifs, recovering the Clausius-Mossotti picture. 

\begin{figure}[h!]
    \centering
    \includegraphics[width=1\linewidth]{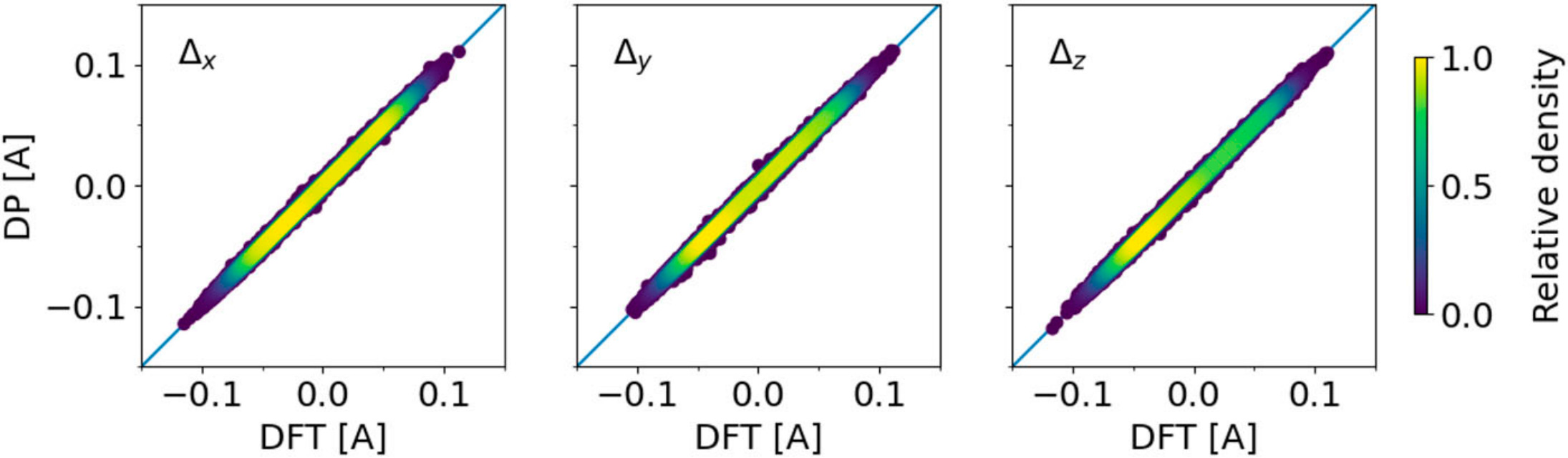}
\caption{{Training accuracy of the DeepWannier model for liquid water. The three panels report the three Cartesian displacements of the Wannier centroids (WCs) relative to their nearest oxygen atom predicted by DW (vertical axis) and by DFT (horizontal axis). Reproduced from Ref.~\cite{Zhang-Weinan2022}, licensed under CC BY 4.0.}}
    \label{fig:wc}
\end{figure}

Models that adopt WC decompositions to represent long-range interactions in machine-learning potentials have been implemented by the DeePMD community in the so-called DeepWannier methods~\cite{zhan+20prb,Zhang-Weinan2022}, which have been recently enhanced with a message-passing framework~\cite{gao2024enhanced}. Such models can be considered as a special flavor of an explicit charge-partitioning scheme that is additionally suited to reproduce the polarization in periodic system. As a key difference with respect to atomic partitioning schemes, the definition of Eq.~\eqref{eq:wannier-pol} implies that the charge values are set as integers, and the model is asked to learn the positions of the (electronic) centers of charge from atomic environments information. In particular, a loss function is defined where a local ML approximation of the WCs is directly compared against reference WC values computed from electronic-structure codes, i.e., 
    $\mathcal{L}({\boldsymbol{\theta}}) = \sum_n\left| \boldsymbol{r}_n({\boldsymbol{\theta}})  -\boldsymbol{r}^\text{ref}_n\right|$.
{An example of the WCs training accuracy is reported in Fig.~\ref{fig:wc}.}
A representation of the electrostatic energy as in Eq.~\eqref{eq:electro} is finally computed by approximating $\rho_Q$ as the sum of spherical Gaussian charges located at nuclear and WCs positions, with the Gaussian width $\sigma$ treated as a tunable parameter. In so doing, the short-range energy $U_\text{SR}$ is here again defined by subtracting from the total energy the WCs approximation of $U_\text{ele}$, and then separately learned with a short-range ML potential model~\cite{Zeng2023}.

A few remarks are needed: the position of a WC can be shifted by choosing a different gauge, i.e., by applying a unitary transformation to the Bloch orbitals used for Wannierization \cite{marzari2012maximally}. It is only when the sum over all WCs is considered, giving the total electronic part of the macroscopic polarization that this gauge freedom is lifted. One should then be aware not to mix datasets where the target WCs were obtained under different prescriptions. Even when a given prescription is used -- such as the Marzari–Vanderbilt maximal localization criterion~\cite{marzari2012maximally} -- to identify a \textit{unique} set of Wannier functions (and hence of WCs), ambiguities may still arise due to the choice of the energy window or, equivalently, of which bands to include or freeze during the Wannierization process.   
Another, more physical, source of ambiguity arises from the possibility of electron transfer between atoms, which undermines the unique assignment of a given set of WCs to specific atomic sites. Ensuring that such situations do not occur is particularly important when partial sums of WCs are computed and associated with a specific chemical moiety. For example, in liquid water, one may choose to learn not all the individual WCs, but rather the Wannier \textit{centroid} of the molecule, defined as the charge barycenter of the WCs corresponding to the O–H bonds and the oxygen lone pairs \cite{Zhang-Weinan2022}. However, if chemical reactions involving electron transfer take place, such a construction ceases to be meaningful. 

DeepWannier has been successfully applied to capture long-range interactions in electronic insulators such as bulk water systems and interfaces. As a special feature of the method, applications have also been demonstrated in the simulation of  the hydrated excess electron~\cite{gao2024enhanced}. In fact, as long as the excess charge distribution remains localized in specific regions of the periodic box, the WC representation can automatically be used to pinpoint the location of the excess electron. Moreover, applications to electrified solid/electrolyte interfaces have demonstrated the capability of the method to make use of the long-range atomic forces to enforce ionic charge electroneutrality in the liquid bulk, thus fully accounting for the effect of perfect ionic screening at thermodynamic equilibrium~\cite{Zhang2024natcomm}, as we will discuss more extensively in Sec.~\ref{sec:discussion}. 

An interesting variation of  {a long-range electrostatic} model compliant with MTP has been recently presented in Ref.~\cite{li-scandolo2026}. Instead of computing the Wannier centers, atomic dipoles $\boldsymbol{d}_i$ are here introduced as auxiliary variables, which are then added to a fixed atomic-charge partitioning $\{q_i^0\}$ when writing the approximation of $U_\text{ele}$. In this case, the total polarization of the system is implicitly defined {similarly to what proposed in Ref.~\cite{Li2025PINN}} as
\begin{equation}
    \boldsymbol{P} = \sum_{i=1}^N \left( q^0_i \boldsymbol{R}_i + \boldsymbol{d}_i \right)\, .
\end{equation}
In line with the learning procedure encountered in implicit atomic-charge models, the atomic dipoles $\boldsymbol{d}_i$ are here predicted through the use of local environment features entering a message-passing NN that is trained on global physical quantities. Importantly, however, the model is not  trained exclusively on electronic energies and forces, but also on Born effective charges (BECs), defined via the matrix $\mathbf{Z}^\star_j = \Omega \,\partial \boldsymbol{P}/\partial \boldsymbol{R}_j$ as the derivative of the macroscopic polarization with respect to atomic displacements. As a result, the model enforces MTP criteria by predicting atomic dipoles, and derived long-range atomic forces, consistently with the topological definition of the polarization in the system. By fixing the atomic charges on each atom, we note that a similar model shares with the Wannier centers models the impossibility of describing macroscopic electronic charge rearrangements and its thus best suited to deal with electronic insulators. Relevant applications of the method have been reported for the longitudinal–transverse optical (LO–TO) phonon splitting in (classical) NaCl solids, a prototypical example characterized by a macroscopic response phenomenon, which is also correctly capture via the DeepWannier model \cite{Zhang-Weinan2022}.  Finally, as we will discuss in more details in Sec.~\ref{sec:finite-field}, it should be noted that the possibility of predicting Born effective charges represents a natural strategy for introducing the interaction with an external electric field.

\section{Electrostatics from nonlocal models}

As formulated in Eq.~\eqref{eq:locality-constraint}, ML approaches that predict an approximation of $\rho_Q$ solely from the local structural information around the atoms of the system cannot reproduce polarization effects that originate from structural changes occurring beyond the model cutoff. 
While this type of physics adds a second-order correction to the charge approximation, there are cases where incorporating this long-range effect is especially important~\cite{Unke2021chemrev}.  For example, a pathological manifestation of electronic polarization is found when considering electronically conductive frameworks such as metallic compounds and elongated aromatic chains. In these cases, the polarization phenomenon can be equivalently interpreted as the manifestation of an electronic charge transfer across macroscopic regions of the system, requiring the ML model to acquire a nonlocal character at inference. 
Importantly, while the electrostatic energy as defined in Eq.~\eqref{eq:electro} is inherently nonlocal due to the presence of the Coulomb potential, nonlocality is here interpreted in the sense of overcoming the constraint of Eq.~\eqref{eq:locality-constraint}. 

\begin{figure}[t!]
    \centering  \includegraphics[width=1\linewidth]{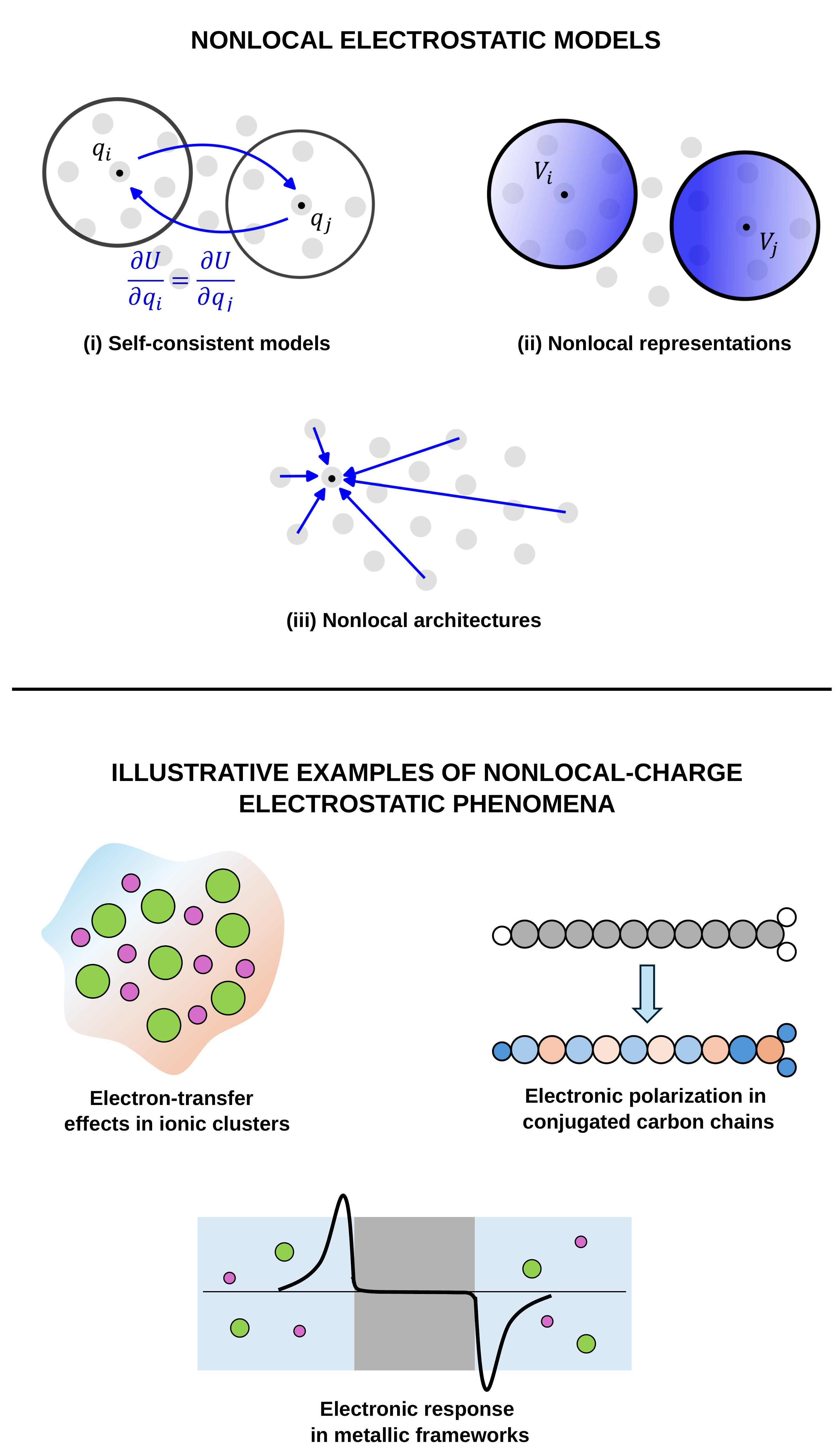}
    \caption{{\textit{Top}: families of nonlocal electrostatic models in atomistic machine learning: (i) atomic charges are self-consistently determined to satisfy a physics-based variational principle, (ii) long-range electrostatic representations built from all the atoms in the system are used as input features of the model, (iii) nonlocal architecture are adopted to enforce an all-to-all communication among the atoms in the system. \textit{Bottom}: Illustrative examples of systems characterized by electronic charge transfers and nonlocal polarization effects.}}
    \label{fig:method-2}
\end{figure}

In this section, we show that three different paradigms can be adopted to endow the model with nonlocality, as depicted in Fig.~\ref{fig:method-2}: (i) the implementation of a self-consistent procedure capable of equilibrating the charge distribution in the system based on some variational criteria, (ii) the incorporation of nonlocal input representations of \textit{all} atomic positions in the system, (iii) the implementation of nonlocal operations within the learning architecture.

\subsection{Nonlocality via self-consistency~\label{sec:QEq}}
Well before the advent of ML potentials, one of the pioneering approaches developed to account for nonlocal charge-transfer effects into classical molecular dynamics simulations consisted in the so-called charge-equilibration (QEq) method~\cite{rappe1991charge,Wilmer2012}. Similarly to atomic partitioning schemes already encountered, this method relies on a  representation of the  charge density on a set of atomic charges $\{q_i\}$. In QEq, however, the atomic charges are not directly fitted from computed quantum data of electrostatic properties, but they are self-consistently optimized to satisfy an (electronic) chemical potential equalization (CPE) principle. Specifically, a second-order expansion of the electronic energy with respect to $q_i$ is first derived in terms of atomic energies, $U_i^0$, electronegativies $\chi^0_i$, and chemical hardness $J^0_{ii}$, which is eventually added to the electrostatic energy associated with the Coulomb interaction $J_{ij}$ between the atomic charges:
\begin{equation}\label{eq:QEq}
    U(\{q_i\}) = \sum_i U_i^0 + \sum_i \chi^0_i q_i + \frac{1}{2} \sum_{ij} \left(J^0_{ij}\delta_{ij}+J_{ij}\right) q_i q_j \, .
\end{equation}
Following the original formulation of Rappe and Goddard~\cite{rappe1991charge}, both $\chi^0_i$ and $J^0_{ii}$ can be approximated from tabulated values ionization potentials (IP) and electronic affinities (EA) of the isolated atoms, while the form of $J_{ij}$ is once again defined as a Coulomb potential between the atoms, damped at small interatomic distances. From the former definition of electronic energy, equalizing the electronic chemical potential of each of the $N$ atoms in the system implies setting the $N-1$ equalities: 
\begin{equation}
     \frac{\partial U(\{q_i\})}{\partial q_1}=\frac{\partial U(\{q_i\})}{\partial q_2}=\ldots=\frac{\partial U(\{q_i\})}{\partial q_N}\, .
\end{equation}
The conservation of the total charge $Q$, i.e., $\sum_i q_i = Q$, finally sets the $N$-th condition needed for solving the QEq problem as a linear system of equations and finding the optimal charge partitioning $\{q_i\}$. We note that an equivalent solution can be found from direct minimization of the electronic energy of Eq.~\eqref{eq:QEq}, by introducing a Lagrange multiplier $\lambda$ for the conservation of the total charge. In this context, the value of $\lambda$ has the role of enforcing the CPE condition by providing an external constraint to the minimization procedure. A simplified illustration of the QEq scheme is reported in Fig.~\ref{fig:method-2}-(i). {A formally developed framework for incorporating self-consistently Coulomb interactions into otherwise local ML interatomic potentials, along with theoretical insights into why QEq–augmented ML potentials can be successful, is presented in Ref.~\cite{thomas2025self}.}

Within QEq, the incorporation of the long-range electrostatics is implicitly achieved via the variational procedure at the cost of a matrix inversion, or explicit minimization of the QEq energy. The use of machine learning in this context has resulted in diverse approaches that make the QEq ingredients dependent on the local atomic environment. The first formulation of a ML-QEq method is the charge equilibration neural network technique~\cite{Ghasemi2015} (CENT), where a NN model is used to provide an environment-dependent approximation of the atomic electronegativities $\chi_i$. These are implicitly optimized via the QEq scheme to reproduce DFT-level electronic energies, i.e.,  $ \mathcal{L}({\boldsymbol{\theta}}) = \left|U(\{\chi_i({\boldsymbol{\theta}})\})-U^\text{ref}\right|$. Moreover, a Gaussian smearing of the atomic charges is adopted to provide a smoother representation of $\rho_Q$ -- a strategy that is used in all modern charge-equilibration methods. By automatically accounting for nonlocal electronic charge rearrangements, CENT achieved unprecedented accuracy in reproducing the energetics of neutral and ionized NaCl clusters. 

\begin{figure}[h!]
    \centering
    \includegraphics[width=0.9\linewidth]{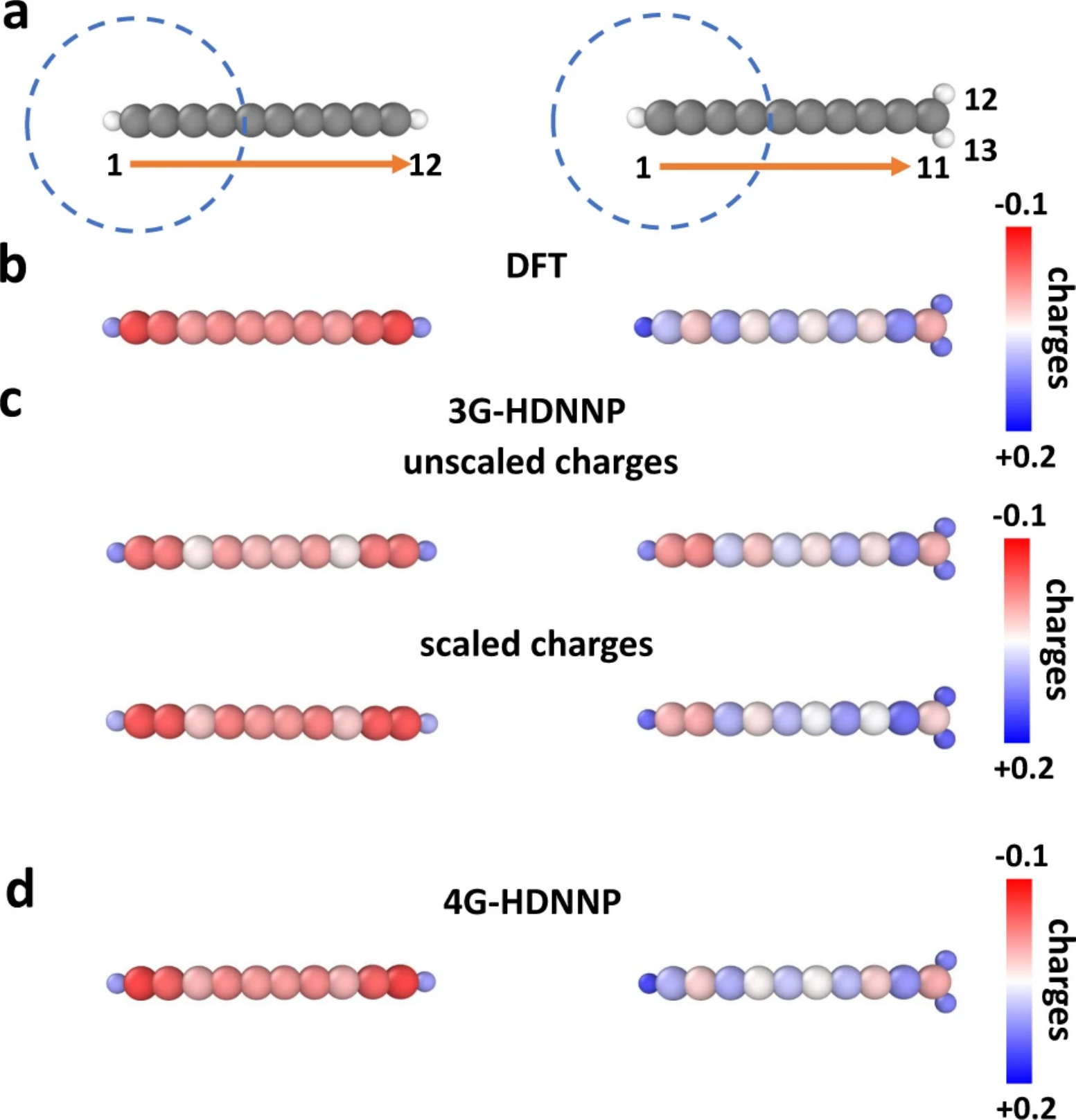}
    \caption{{(a) DFT-optimized structures of C10H2 (left) and C10H (right) with atom IDs. Carbon and hydrogen atoms are colored in gray and white, respectively. The dashed circle shows the cutoff radius of the left carbon atom defining its chemical environment. (b) shows the atomic partial charges obtained from DFT. The unscaled and scaled 3G-HDNNP charges are displayed in (c), while the 4G-HDNNP charges are shown in (d). Reproduced in Ref.~\cite{Ko2021}, licensed under CC BY 4.0.}}
    \label{fig:carbon-chain}
\end{figure}

Successful evolutions of this idea have been developed in the fourth-generation of high-dimensional neural-network potentials, i.e., 4G-HDNNP~\cite{Ko2021,Gubler2024,Kocer2025}. Unlike CENT -- where the charges obtained from the QEq scheme serve merely as auxiliary variables -- 4G-HDNNP fits the electronegativities to reproduce reference Hirshfeld charges computed from DFT. {Such an approach, aiming to reproduce Hirshfeld charges, is also shared by Ref.~\cite{maruf2025learning}.}
Similarly to the learning rationale adopted in the 3G-HDNNP model, 4G-HDNNP separates a short-range contribution of the electronic energy $U_\text{SR}$, which is learned via a separate neural network using the equilibrated $q_i$ as additional descriptor parameters of the local atomic environment. 
Variational optimizations of 4G-HDNNP have been later proposed that streamline the learning procedure thanks to a multi-targeted loss function where all expansion coefficients of the QEq energy in Eq.~\eqref{eq:QEq} are simultaneously represented by via a single-pass NN prediction~\cite{Shaidu2024}. Several applications and model validations of QEs schemes appear in the literature: The example of a diatomic gold cluster (Au$_2$) on the (001) surface of a MgO slab, doped with aluminum atoms introduced in the inner layers of the slab, i.e.~several \AA~away from the Au$_2$ cluster, demonstrates the critical role of QEq models in accurately capturing the long-range charge transfer that alters the electronic structure and energetic stability of the system, thereby driving the transition between wetting and non-wetting adsorption geometries of the Au$_2$ cluster~\cite{Ko2021,Gubler2024,Shaidu2024}. Another recent example is given by molecular dynamics simulations of a FeCl$_3$-water system, demonstrating that global charge equilibration is essential for accurately predicting iron oxidation states (Fe$^{2+}$ or Fe$^{3+}$), defined by distant chloride counter ions~\cite{Kocer2025}.
Additionally, benchmarks have been carried out on pairs of systems with different charge states, such as neutral and protonated carbon chains {(Fig.~\ref{fig:carbon-chain})}, oppositely charged silver clusters, and cationic sodium chloride clusters of varying stoichiometry, demonstrating the model's ability to correctly describe finite systems with arbitrary net charges~\cite{Shaidu2024}.
Researchers' effort has also focused on obtaining more efficient solutions of the linear system in the expensive charge equilibration step~\cite{Gubler2024,Kocer2025}.

As a shared aspect of the discussion carried out in Sec.~\ref{sec:local-models}, the inherent arbitrariness of Hirshfeld partitioning schemes can be bypassed by training the QEq model on global learning targets.
While the CENT method already achieves this goal through the learning of electronic energies, a kernel-based version of the QEq method (kQEq) has been proposed where the electronegativies of the model are found to indirectly reproduce computed DFT data for the molecular dipole moment~\cite{Staacke2022}. The model then relies on a loss function similar to that reported in Eq.~\eqref{eq:dipole}. Interestingly, a direct accuracy comparison against the MuML model~\cite{veit+20jcp}  already discussed in Sec.~\ref{sec:local-models}  was performed when predicting the dipole moment of increasingly long zwitterionic molecules that include aromatic carbon chains along the molecular backbone. It was found that the incorporation of self-consistency into the model is beneficial for better distributing the charge on the carbon atoms, although the dipole moment is, in this example, mostly determined by the permanent opposite charges located at the molecular ends.
Another recent development in the context of graph neural networks  and message passing  is CELLI \cite{fuchs2025learning}, which equips the message-passing backbone with a QEq solver: local message-passing features from the GNN determine environment-dependent electronegativities, and a global charge-equilibration step then produces self-consistent charges that reintroduce nonlocal long-range polarization effects beyond what finite-cutoff message passing can capture. Validation on benchmark systems -- including isolated gas-phase clusters and clusters adsorbed on extended supports (e.g., the already described gold dimer on an MgO(001) surface) -- demonstrated state-of-the-art accuracy for capturing nonlocal interactions.
{Polarizable charge-equilibration (PQEq) stratergies have also been developed: Ref.~\cite{gao2025foundation} introduce a ML potential that couples an equivariant GNN short-range model with an explicit PQEq long-range electrostatics module, yielding self-consistent polarization and auxiliary partial charges.}

A subtle but important issue in constructing ML force fields is whether the resulting model is \emph{conservative}, i.e.,\ whether the forces derive from a well-defined scalar potential energy~\cite{Bigi2025}. Charge–equilibration schemes naturally satisfy this requirement as long as the atomic charges are obtained by the minimization of an explicit energy functional and the corresponding solution is differentiable. In this case, although the \textit{optimized} charges $q_i^\ast(\{\boldsymbol{R}_j\})$ depend implicitly on all atomic positions, these dependencies do not contribute to the forces. Specifically, we have from the chain rule,
\begin{equation}
\frac{\mathrm{d} U(\{\boldsymbol{R}_j\},\{q_i^\ast\})}{\mathrm{d} \boldsymbol{R}_k}
= \left.\frac{\partial U}{\partial \boldsymbol{R}_k}\right|_{q^\ast}
+ \sum_{i=1}^N \frac{\partial U}{\partial q_i}\,\frac{\partial q_i^\ast}{\partial \boldsymbol{R}_k}\, ,
\end{equation}
and the second term vanishes     because $\partial U/\partial q_i = \lambda$ for any atom $i$, at convergence, implying
\begin{equation}
    \sum_{i=1}^N \frac{\partial U}{\partial q_i}\,\frac{\partial q_i^\ast}{\partial \boldsymbol{R}_k} = \lambda \frac{\partial \sum_{i=1}^N q_i^\ast}{\partial \boldsymbol{R}_k} = \lambda \frac{\partial Q}{\partial \boldsymbol{R}_k} = 0\, .
\end{equation}
Consequently, differentiable QEq-type models naturally yield conservative force fields. By contrast, machine-learning approaches that \emph{predict} charges directly, without deriving them from an underlying variational principle, do not automatically guarantee conservativity, since no energy-minim\-ization condition enforces the cancellation of these additional terms. In these latter cases, the derivatives of the atomic charges with respect to the atomic positions must therefore be explicitly included as  additional contributions to the atomic forces.

\begin{figure}[h!]
    \centering
    \includegraphics[width=1\linewidth]{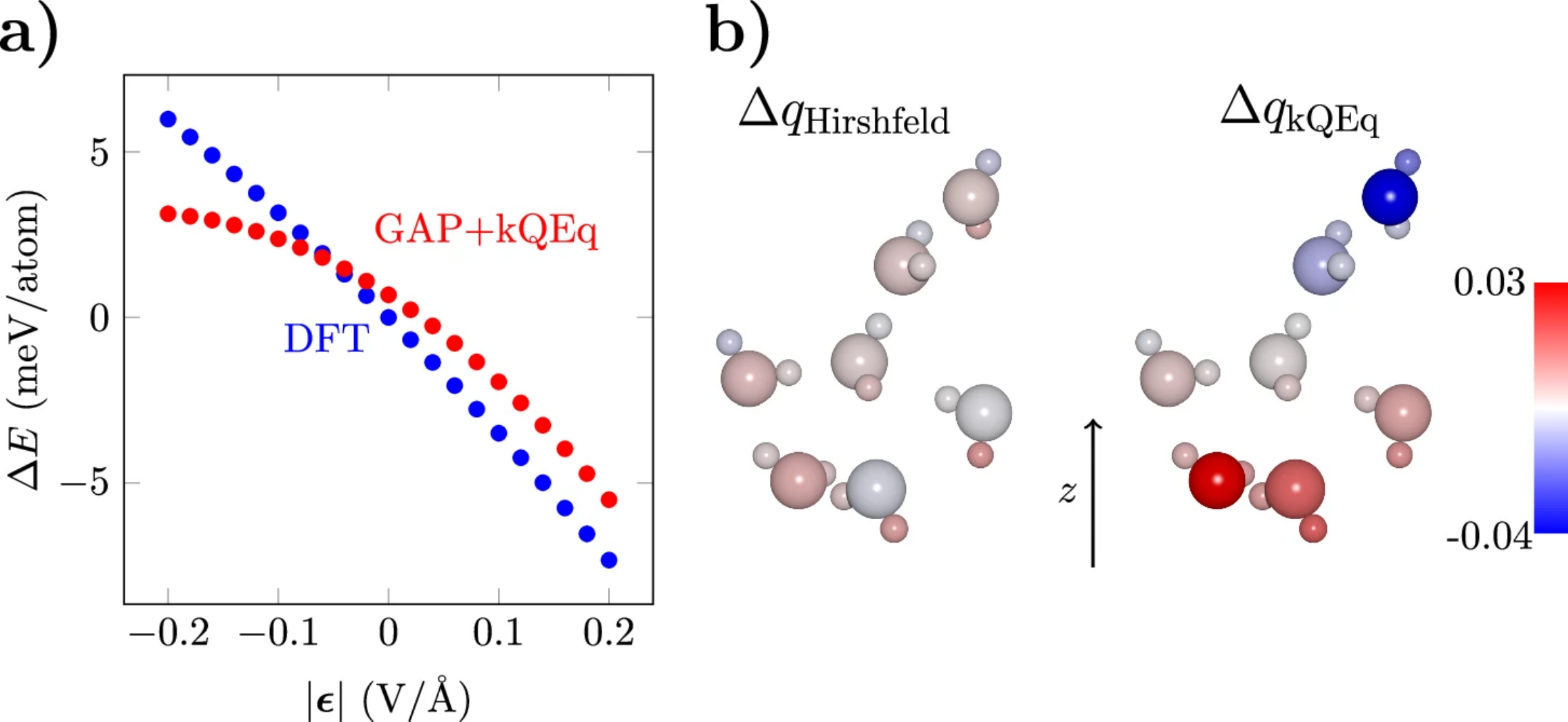}
    \caption{{(a) Energetic response of DFT and ML-QEq for a neutral water cluster subjected to external electric fields of varying magnitudes along a fixed direction. Energies are shifted so that the DFT energy is zero at zero field. (b) Difference in atomic charges obtained from Hirshfeld partitioning (DFT) and ML-QEq under external fields of -0.2 V/{\AA} and 0.2 V/{\AA}. Reproduced from Ref.~\cite{Vondrak2025}, licensed under CC BY 4.0.}}
    \label{fig:water-cluster}
\end{figure}

Overall, QEq methods provide an excellent compromise between the locality of the ML structural representation and the nonlocal character of the physical target. However, the minimization procedure required to obtain the self-consistent solution of the atomic charges adds a computational overhead at inference that can become problematic when indefinitely extending the system size. Furthermore, it should be noted that their high accuracy is mostly restricted in the domain of electronically conductive materials, where a chemical-potential equalization principle can be rigorously  defined. This structural limitation of the model has been recently well characterized in Ref.~\cite{Vondrak2025}, where it is shown how the equilibration procedure can lead to a diffuse overpolarization effect when exposing molecular water clusters to an applied electric field -- rather than limiting the polarization at the level of the individual molecules, as expected in electronically insulating frameworks {(Fig.~\ref{fig:water-cluster})}. In this context, QEq methods represent a complementary approach to Wannier centers models, so that a suitable combination of the two could possibly be developed in the future to bridge the gap between the two extreme case scenarios associated with purely conductive and insulating materials.

Of course, ML models based on QEq are not the only way one could incorporate nonlocal effects via a self-consistent equilibration procedure.
In fact, a different (still largely unexplored) possibility is that of introducing self-consistency as an integral part of the learning algorithm. A notable example of this was presented in the self-consistent field neural network (SCFNN) model~\cite{Gao2022}. In this case, the long-range atomic forces in the system are predicted based on the evaluation of an electric-field descriptor $\boldsymbol{E}(\boldsymbol{r})$ that is added to the local representation of the atomic coordinates. In turn, this field is self-consistently optimized to account for far-field effects using a linear-response formalism that includes an approximation of the physical charge density $\rho_Q$:
\begin{equation}
    \boldsymbol{E}(\boldsymbol{r}) = \boldsymbol{E}_\text{ext}(\boldsymbol{r}) - \int \mathrm{d}\boldsymbol{r}'\, \rho_Q(\boldsymbol{r}')\, \nabla_{\boldsymbol{r}} v_\text{C}(|\boldsymbol{r}'-\boldsymbol{r}|)\, ,
\end{equation}
where $\boldsymbol{E}_\text{ext}$ is a possibly applied external field, and $v_\text{C}$ is the Coulomb potential. In particular, $\rho_Q$ is computed from the sum of nuclear charges and Wannier centers, whose positions $\boldsymbol{r}_n$ are self-consistently optimized through a separate neural network using the solution for $\boldsymbol{E}(\boldsymbol{r})$ as a descriptor parameter for the subsequent NN iteration. Crucially, in order for the NN prediction to be sensitive to fields of different strengths used as input, the model is trained on a reference dataset of Wannier centers computed under external electric fields of varying intensities. In practice, SCFNN adopts a range separation strategy where conventional local NN models are used to learn the short-range contribution to atomic forces and Wannier centers, implying that the self-consistent procedure only optimizes the WC variations $\Delta \boldsymbol{r}_n$ induced by long-range potential components. This model has proven to reproduce the $k\to 0$ (dielectric) limit of the longitudinal component of the dipole correlation function in bulk liquid water, thus providing a clear signature for the inclusion of long-range interactions. However, we recognize how a similar property is not directly related to the manifestation of nonlocal polarization effects, so that other local electrostatic models could in principle be used to achieve the same result. In this context, the use of similar methods to the prediction of nonlocal electronic polarizations should be explored in the future to assess the added value of integrating self-consistency within the learning architecture.

\subsection{Nonlocality via nonlocal representations~\label{sec:nonlocal-rep}}

Beyond a self-consistent treatment of the electrostatic energy, a seemingly natural way to include nonlocal physical effects into ML models is that of adopting \textit{nonlocal} representations of the atomic structure 
-- see Fig.~\ref{fig:method-2}-(ii). In fact, as long as all atomic coordinates in the system are simultaneously used in input, the model possesses, in principle, all the information required to approximate the physical target, including nonlocal structural correlations.
On a general level, a nontrivial aspect in the use of nonlocal representations is that of preserving the \textit{size-extensive} character required to achieve linear-scaling predictions with the number of atoms in the system, while also enabling a \textit{size-consistent} description of long-range interactions. A size-consistent model should be able to reproduce the dissociated (noninteracting) limit between sub-portions of the system that are brought at a virtually infinite mutual distance~\cite{Pople1976}, i.e., $U(R_{AB}\to\infty) = U(A)+U(B)$ -- a condition that can be difficult to satisfy even in DFT methods due to the well-known electronic delocalization error of common density-functional approximations~\cite{Zheng2012}. Local ML models naturally satisfy size-extensivity, but they cannot generally guarantee size-consistency when the interactions in the system extend beyond the local environment cutoff. 

Straightforward implementations of nonlocal ML representations are global descriptors of the atomic structure that get rid altogether of the notion of local atomic environments. A paradigmatic example, although not strictly applied to the prediction of long-range properties, is given by Coulomb matrices~\cite{rupp+12prl}.  Here, a molecular matrix $\mathbf{M}$ of the form 
\begin{equation}\label{eq:coulomb-matrix}
    M_{ij} = \begin{cases}
        0.5 Z_i^{2.4}\, , \quad i=j\\\frac{Z_iZ_j}{|\boldsymbol{R}_i-\boldsymbol{R}_j|}\, , \quad i\ne j 
    
    \end{cases}
\end{equation}
is first defined, with $Z_i$ the nuclear charges. Upon diagonalization, the eigenvalues of $\mathbf{M}$ are  used to construct an input-vector representation for a kernel-based ML model. 
This approach has proven to be successful in predicting the atomization energy of heterogeneous datasets of small organic molecules~\cite{rupp+12prl,Hansen2013}.
Because of the lack of inherent additivity, however, ML models  based on  global descriptors are generally neither size-extensive nor size-consistent, so that their accuracy is mostly limited in learning molecular properties associated with datasets covering small size variations. In this context, it has been pointed out how  size-extensive predictions can be implicitly recovered when using global descriptors within a linear regression framework, providing the model has learned how a given property scales with the number of atoms in the system~\cite{Jung2020}. This has, in particular, been proven in the context of atomization energies when adopting many-body tensor representations (MBTR~\cite{Huo2022}) as global descriptors. Because of the imposed linearity in the model's inputs, size-consistency could also be attained in this case by exploiting the separability of the global descriptor in local fragments. This is realized,  for instance, from the block diagonal character of the Coulomb matrix of Eq.~\eqref{eq:coulomb-matrix} when considering two noninteracting portions of the system. However, similar approaches have, to our knowledge, so far not been tested in the context of reproducing long-range interactions, highlighting the need to explore this direction.

Nonlocal representations of the atomic structure that are both size-extensive and size-consistent would naturally require to conserve the additivity derived from atom-centered descriptors while effectively incorporating long-range information that extends indefinitely beyond the local environment cutoff. Early developments consistent with this criterion can be found in the long-distance equivariants (LODE) model~\cite{gris-ceri19jcp}. In LODE, the notion of atomic environment is lifted by evaluating a fictitious approximation of the electrostatic potential, $V(\boldsymbol{r})$, rather than a local representation of the atomic positions, within the spherical window of radius $r_\text{cut}$ centered on a given atom~$i$. 
Specifically, $V$ is first derived by applying the Coulomb integral operator on a smeared Gaussian-based representation, $\rho_\sigma(\boldsymbol{r})$, of \textit{all} atomic positions in the system:
\begin{equation}\label{eq:lode}
    V(\boldsymbol{r}) = \int \mathrm{d}\boldsymbol{r}'\, \frac{\rho_\sigma(\boldsymbol{r'})}{|\boldsymbol{r}'-\boldsymbol{r}|} = \int \mathrm{d}\boldsymbol{r}'\, \frac{\sum_{j=1}^N \exp\left(-\frac{|\boldsymbol{r}'-\boldsymbol{R}_j|^2}{2\sigma^2}\right)}{|\boldsymbol{r}'-\boldsymbol{r}|}\, ,
\end{equation}
where, in practice, the sum is restricted to the atoms of a given chemical species, so that $V$ becomes species-dependent. 
For each atom $i$ in the system, this descriptor is then centered at the atomic positions $\boldsymbol{R}_i$ and a spherical cutoff function $f_\text{cut}$ is finally applied to ensure that the potential field is only evaluated locally around the atoms, i.e.,
\begin{equation}
    V_i(\boldsymbol{r}) = f_\text{cut}(|\boldsymbol{r}-\boldsymbol{R}_i|)\, V(\boldsymbol{r}-\boldsymbol{R}_i)\, .
\end{equation}
Crucially, because of the pathologically slow decay of the Coulomb potential, each atomic environment remains sensitive to structural variations \textit{everywhere} in the system, i.e., $\partial V_i/\partial \boldsymbol{R}_k\ne 0$ for any atom $k$ (even beyond the radial cutoff $r_\text{cut}$) thereby turning $V_i$ into a genuinely nonlocal atom-centered descriptor. Subsequent equivariant combinations of $V_i$ with possibly different types of atom-centered descriptors can be performed to ensure that the final LODE representation satisfies prescribed symmetry criteria. We note that implementations of LODE to both molecular and periodic systems can be achieved from real and reciprocal space formulations of Eq.~\eqref{eq:lode}, sharing with other methods the possibility of making Ewald-like evaluations computationally efficient thanks to the adoption of particle-mesh algorithms~\cite{Loche2025}. 

A different type of atom-centered long-range descriptor, which does not rely on the explicit calculation of the Coulomb potential, has more recently been proposed~\cite{Faller2024}. Here,  the atom-density $\rho_\sigma$ is evaluated around each atom through  projection on exponential radial functions designed to accurately reproduce a $1/r$ decay via specific quadrature rules~\cite{Hackbusch2019}. In so doing, the model does not apply an explicit cutoff function, but rather let the decay rate of the basis functions determine the spatial extent of each atom-centered descriptor. 

As a marked difference with respect to electrostatic-energy models that rely on charge approximations, the fact that the nonlocal character is encoded in the definition of the structural descriptor implies that this family of methods is not limited to predict a prescribed long-range property. This inherent flexibility can, however, lead to ambiguities in determining which type of long-range physics is actually captured during the training procedure. 
In the case of LODE, one would be interested in establishing up to which extent the set of $\{V_i\}$ can be used to incorporate nonlocal physical information within a (possibly nonlinear) ML functional approximation of the electrostatic energy expressed as a sum over atomic contributions, i.e., $
U_\text{ele}[\{V_i\}] = \sum_{i=1}^N F_{\boldsymbol{\theta}}[V_i] $. To see this, let us explicitly write the derivative of $U_\text{ele}$ with respect to an arbitrary atomic displacement; after applying the chain rule, we obtain:
\begin{equation}\label{eq:lode-chain}\begin{split}
\frac{\partial U_\text{ele}[\{V_i\}]}{\partial \boldsymbol{R}_k } = &\sum_{i=1}^N \int \mathrm{d}\boldsymbol{r}\, \frac{\delta F_{\boldsymbol{\theta}}[V_i]}{\delta V_i(\boldsymbol{r})} \, f_\text{cut}(|\boldsymbol{r}-\boldsymbol{R}_i|) \\& \times\int \mathrm{d}\boldsymbol{r}'\, \frac{1}{|\boldsymbol{r}'-\boldsymbol{r}-\boldsymbol{R}_i|} \frac{\partial \rho_\sigma(\boldsymbol{r}')}{\partial\boldsymbol{R}_k}\, .
\end{split}\end{equation}
This expression suggests that   displacements of any atom $k$ is propagated to the  environment of atom $i$ via the Coulomb potential, thus establishing a nonlocal coupling between the fixed Gaussian atom-density $\rho_\sigma$ and the functional derivative $\delta F_{\boldsymbol{\theta}}/\delta V_i$. In the simplest case where $F_{\boldsymbol{\theta}}[V_i]$ is a linear functional of $V_i$, $\delta F_{\boldsymbol{\theta}}/\delta V_i$ is a local quantity, implying that the nonlocality of the model is only used to provide an agnostic approximation of the Hartree integral of Eq.~\eqref{eq:electro}.  Therefore, linear LODE models of the  electrostatic energy appear as formally compliant with the charge locality criterion stated in Eq.~\eqref{eq:locality-constraint}.

A relevant case scenario consistent with the previous discussion is found when considering a linear approximation of the electrostatic energy built from LODE features computed from the product of atom-density ($\rho_i$) and potential ($V_i$) descriptors. Upon algebraic manipulations, this model yields exact asymptotic relations that match the physics of multipoles interactions~\cite{grisafi2021cs}:
\begin{equation}\label{eq:multipoles}
      U_\text{ele}[ \{V_i\}] =  \sum_{i=1}^N\sum_{l=0}^{l_\text{max}} \sum_{m=-l}^l M^{lm}_i({\boldsymbol{\theta}}) \int^\infty_{r_\text{cut}} \mathrm{d}r\, \frac{\rho^{lm}_\sigma(r)}{r^{l+1}} \, , 
\end{equation}
with $l_\text{max}$ determining the maximum angular momentum included in the multipolar expansion.
Here, the term asymptotic is interpreted as evaluating the interaction between (fictitious) atomic multipoles, $M^{lm}_i$, implicitly found by the model, and the far-field atom density $\rho_\sigma$ located \textit{beyond} $r_\text{cut}$.
Notably, Eq.~\eqref{eq:multipoles} appears as a multipolar generalization of implicit atomic-charge models already encountered in Sec.~\ref{sec:local-models}. 

\begin{figure}[t!]
    \centering
    \includegraphics[width=0.75\linewidth]{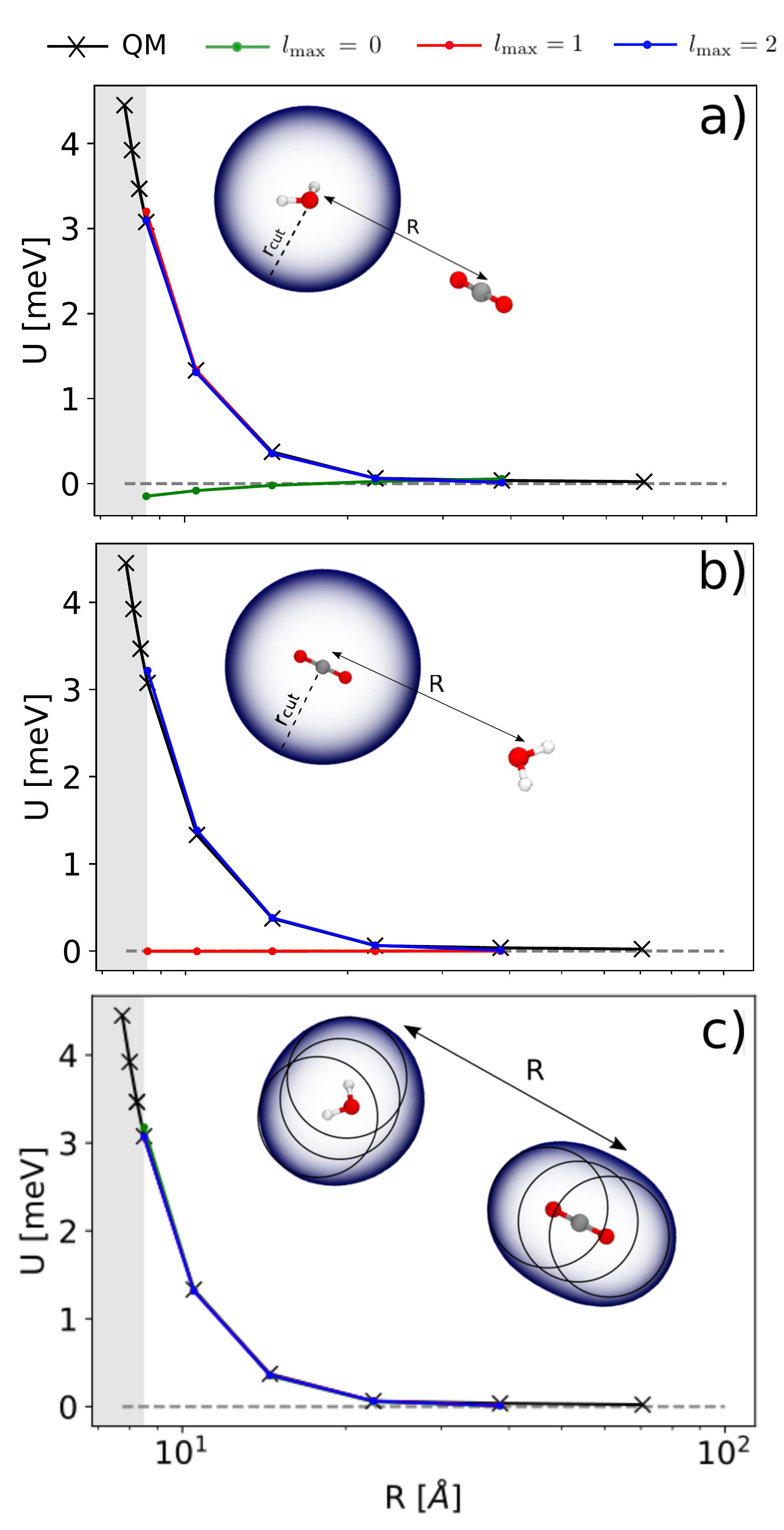}
    \caption{{Interaction energy for a given configuration of H$_2$O and CO$_2$ at different angular cutoffs $l_\text{max}$. Top, middle and bottom panels show the results of the asymptotic extrapolation when centering the representation on the oxygen atom of H$_2$O, the carbon atom of CO$_2$, and all the atoms of the system, respectively. Reproduced from Ref.~\cite{grisafi2021cs}, licensed under CC BY-NC 3.0.}}
    \label{fig:lode}
\end{figure}

Linear LODE models have indeed proven to be especially suitable in describing long-range interactions between local charge distributions, by achieving systematic convergence of the electrostatic energy in datasets made of random gases of point-charges~\cite{gris-ceri19jcp}, as well as by retrieving the exact asymptotic behavior of the binding energy between charged and polar molecules, where the multipolar expansion previously reported can be naturally exploited~\cite{grisafi2021cs}. {For example, it is instructive to consider the idealized case of a pure dipole-quadrupole interaction associated with rigid CO$_2$ and H$_2$O molecules at an arbitrarily large mutual distance (Fig.~\ref{fig:lode}). In this case, one can prove that the exact long-range asymptotics can be retrieved from Eq.~\eqref{eq:multipoles} by increasing the angular order $l$ up to the leading multipole of the given molecule used as center of the LODE representation, i.e., the dipole ($l=1$) for H$_2$O and the quadrupole ($l=2$) for CO$_2$. Conversely, truncating the expansion at $l=0$ is enough in this example when using all atoms in the system as expansion centers, thereby mimicking a simple atomic charge model.}

Similarly to implicit charge models, the statistical determination of long-range correlations in the system make linear LODE models also capable of learning, to some extent, induction and dispersion interactions that are not strictly interpretable in terms of ``permanent'' local charges~\cite{grisafi2021cs}. On this front, we mention how direct extensions of LODE ideas implementing different algebraic decays than the Coulomb potential in Eq.~\eqref{eq:lode} have also been proposed in order to endow the model with other types of long-range decays~\cite{Huguenin-Dumittan2023}. 

\begin{figure}[t!]
    \centering
    \includegraphics[width=1.0\linewidth]{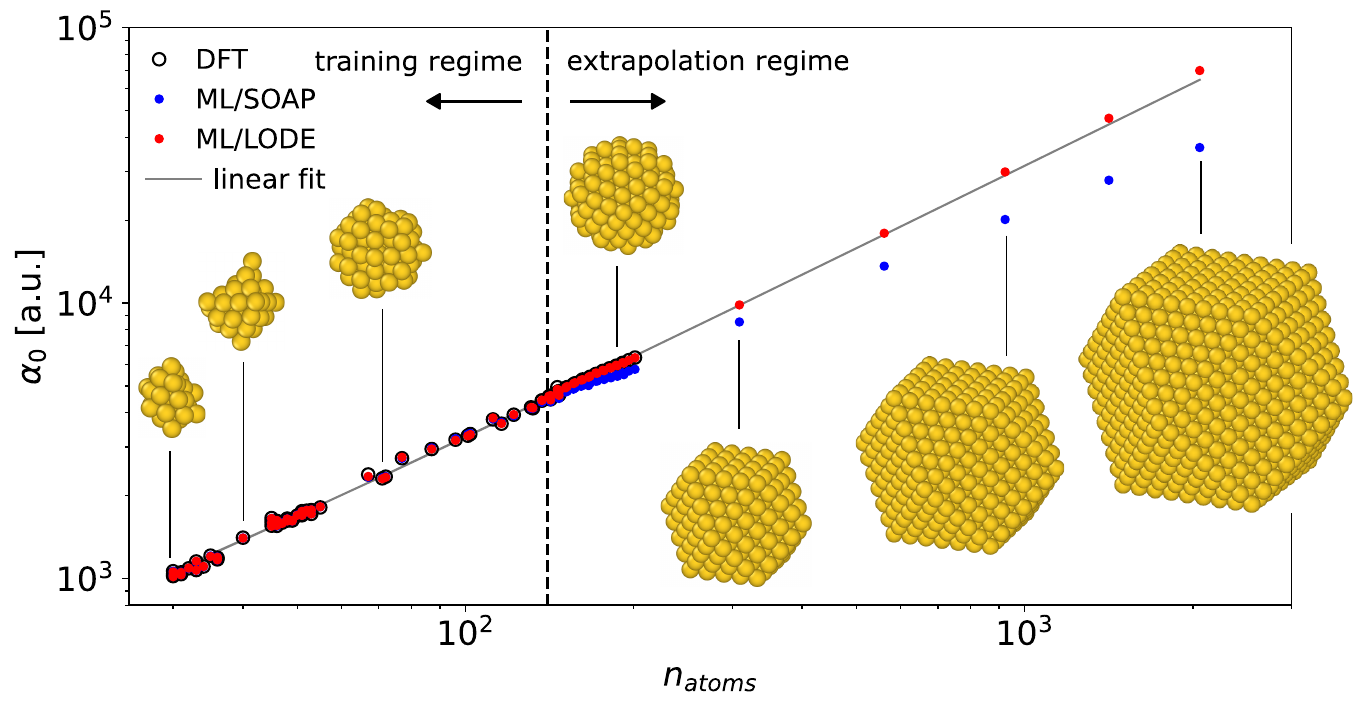}
    \caption{{Isotropic components ($\alpha_0$) of the polarizability tensors of 90 gold nanoparticles of increasing size, as computed from the associated electron density response $\partial \rho_e(\boldsymbol{r})/\partial \boldsymbol{E}$. Empty circles: DFT reference. Red dots: ML prediction. The  dashed line separates the domain of sizes used for training the vector-field model ($N<140$) from the large-scale domain of sizes used to validate the model under extrapolation conditions ($N>140$). Reproduced from Ref.~\cite{Rossi2025}. Copyright 2025 American Chemical Society.}}
    \label{fig:Au-nano}
\end{figure}

Beyond a statistical description of long-range interactions, it is not clear whether the LODE construction could encode enough flexibility to capture the effect of nonlocal polarization phenomena when used to provide a data-driven approximation of the electrostatic energy. While the use of nonlinear ML functionals $F_{\boldsymbol{\theta}}[V_i]$ could be interpreted as going beyond the locality criterion stated in Eq.~\eqref{eq:locality-constraint}, this possibility has not yet seriously been explored and will require a more thorough investigation.  Nonetheless, it should be noted that the explicit prediction of polarization phenomena can alternatively be achieved by LODE via direct learning of the  electronic \textit{charge density}. When adopting an atom-centered decomposition of $\rho_e(\boldsymbol{r})$ analogous to what reported in Eq.~\eqref{eq:RI}, for instance, one can explicitly represent the variation of the electron density components $c^\nu_i$ induced by far-field structural changes as
\begin{equation}
    \frac{\partial c^\nu_i}{\partial \boldsymbol{R}_k} =  \frac{\partial F_{\boldsymbol{\theta}}[V_i]}{\partial \boldsymbol{R}_k} = \int \mathrm{d}\boldsymbol{r}\, \frac{\delta F_{\boldsymbol{\theta}}[V_i]}{\delta V_i(\boldsymbol{r})}\, \frac{\partial V_i(\boldsymbol{r})}{\partial \boldsymbol{R}_k}\, ,
\end{equation}
therefore overcoming the charge-locality criterion of Eq.~\eqref{eq:locality-constraint} by construction. 
Following this approach, it was found that the incorporation of LODE features into electron-density models could accurately reproduce electronic polarization effects in metallic slabs of increasing thickness~\cite{grisafi2023prm}, as well as the static electric-field response, $\partial \rho_e(\boldsymbol{r})/\partial \boldsymbol{E}$, in metallic nanoparticles of increasing diameter~\cite{Rossi2025}, demonstrating the ability to learn nonlocal charge variations across a broad range of system sizes {(Fig.~\ref{fig:Au-nano})}. In perspective, similar strategies could provide a rigorous approach to facilitate the inclusion of physically distinct contributions to the long-range electrostatics -- namely associated with permanent and induction terms -- by distributing the learning burden between an explicit charge-density model and an implicit description of long-range polarization.

{Recent evolutions of LODE ideas that additionally introduce an iterative procedure to mimic induction effects (without enforcing full self-consistency) have recently been proposed in the MACE-POLAR-1 foundation model~\cite{Batatia2026}. In particular, the model adopts a proxy of the spin-charge density written following a linear atom-centered expansion similar to Eq.~\eqref{eq:RI}. The resulting atomic-multipole coefficients are then used as learnable auxiliary variables of the model to build long-range electrostatic features following the LODE construction, which are then used to define the atomic multipoles for the subsequent iteration. Additionally, the model equilibrates the global charge $Q$ following the same Fukui-based strategy~\cite{Zubatyuk2021} already reported in Eq.~\eqref{eq:fukui}, thereby accounting for ionic charge states. Importantly, the model is only trained on energies and forces adopting a conventional decomposition of the electronic energy in a local and electrostatic contribution, where the former is learned via the MACE model~\cite{Batatia2022}, while the latter is computed via the Hartree-energy expression defined from the learned atomic multipoles. In this optic, MACE-POLAR-1 realizes the natural integration between nonlocal structural representations and the implicit atomic-charge models discussed in Sec.~\ref{sec:implicit}, which is further refined by endowing the inference procedure with an iterative charge update. Upon training on $\sim10^8$ hybrid DFT calculations, MACE-POLAR-1 has shown to yield chemical accuracy in a wide variety of contexts, achieving fourfold improvement over short-ranged models on
protein-ligand interactions, as well as an improved description of ions and redox reactions of transition metals in solution~\cite{Batatia2026}. In this context, a systematic analysis that disentangles permanent and induction contributions to the system's electrostatics would be highly valuable to quantify the role of the implemented iterative procedure on the overall predictive performance.}

\subsection{Nonlocality via nonlocal architectures}

The previous discussion exemplifies how ML models can encode nonlocal effects directly through the nonlocality of the input representation. Different literature examples exist that do not directly rely on nonlocal structural features, but rather integrate nonlocal operations within the learning architecture.
In contrast to descriptors such as Coulomb matrices or LODE features, which are constructed deterministically through predefined functional forms, these models rely on learned representations whose structure emerges during training. In these architectures, the atomic features are not fixed by a closed-form formula, but are produced through parametrized message-passing, attention, or mesh-based operations whose weights are optimized jointly with the final target, i.e., the potential energy surface. Although these models typically start from local environments, the successive learned transformations can encode information that reflects the global configuration of the system, enabling the network to capture long-range couplings without requiring an explicitly handcrafted nonlocal descriptor, as schematized in Fig.~\ref{fig:method-2}-(iii).

For example, SpookyNet~\cite{Unke2021} introduces nonlocality into neural network potentials by incorporating nonlocal structural information through a self-attention mechanism that operates over all atoms in the system. As such, the model allows each atom to \textit{attend} to every other atom in the molecule or material, effectively using a context window encompassing the entire atomic set. In practice, each atom's representation is updated through query-key-value interactions with all other atoms, returning attention weights that reflect the learned importance of possibly distant atoms. This global attention captures long-range correlations and collective electronic effects that are otherwise inaccessible to purely local 
charge models. 
Importantly, SpookyNet adopts the FAVOR+ approximation \cite{choromanski2020rethinking}, which reduces the computational scaling of the attention operation from the quadratic $\mathcal{O}(N^2)$ of standard self-attention to linear $\mathcal{O}(N)$, making the inclusion of full-system nonlocality computationally practical.

\begin{figure}[t!]
    \centering
    \includegraphics[width=1\linewidth]{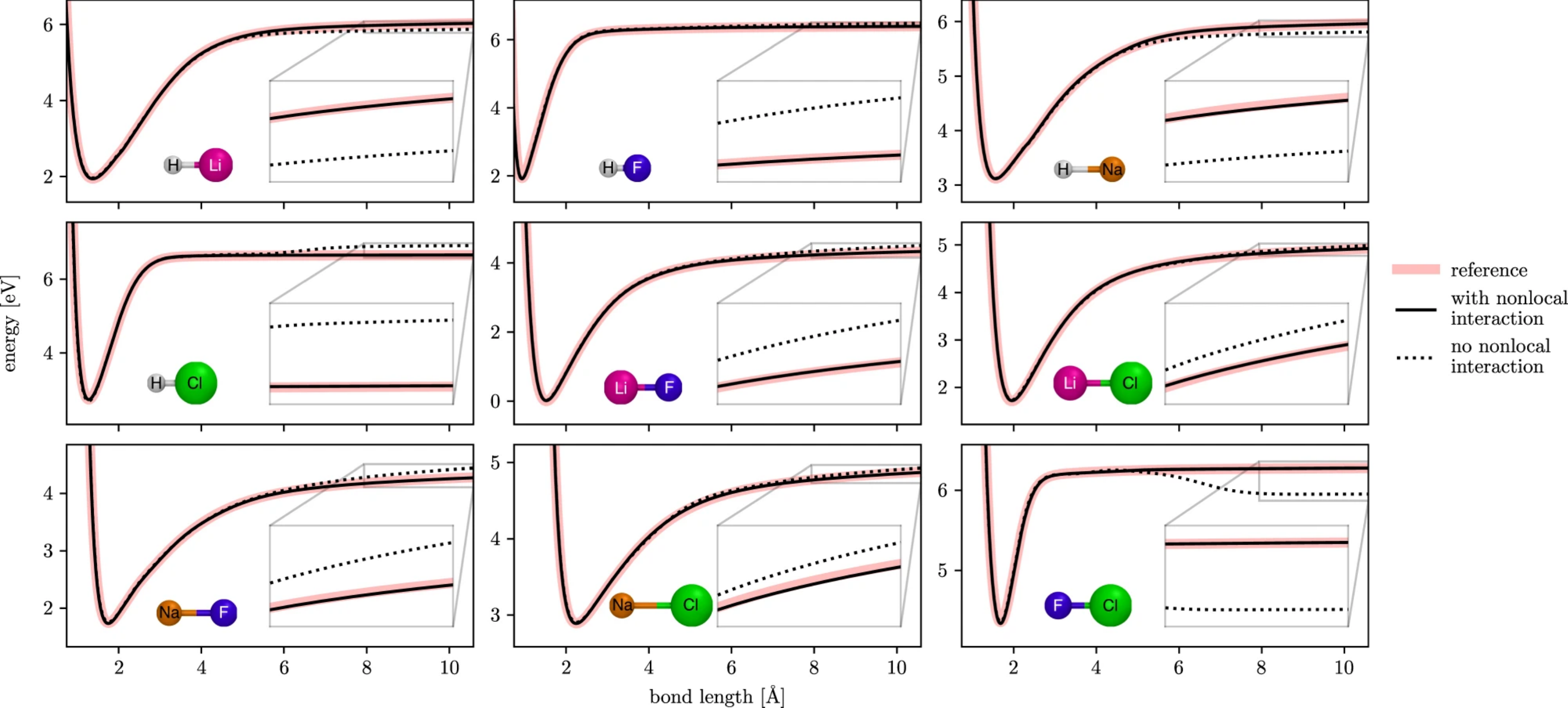}
    \caption{{Dissociation curves of different diatomic molecules predicted by SpookyNet with/without nonlocal interactions. Although long-range electrostatics are included in both cases, the model without nonlocal interactions fails to reproduce the asymptotic tails of the reference calculation at large interatomic separations. Individual panels show the dissociation curves for different species. From top-left to bottom/right: HLi, HF, HNa, HCl, LiF, LiCl, NaF, NaCl, and FCl. Reproduced from Ref.~\cite{Unke2021}, licensed under CC~BY~4.0.}}
    \label{fig:spooky}
\end{figure}

{A relevant application of SpookyNet showing the importance of nonlocal interactions for the dissociation of diatomic molecules is reported in Fig.~\ref{fig:spooky}.} Along the same lines, So3krates~\cite{Thorben2022} introduces nonlocality through an attention-based mechanism. Nonetheless, rather than employing full all-to-all self-attention over atomic embeddings, So3krates augments a local, cutoff-based equivariant message-passing scheme with an additional aggregation performed in spherical-harmonic coordinate (SPHC) feature space. In practice, atomic spherical-harmonic projected coefficients are updated by two neighbor sums: a conventional real-space sum over Euclidean neighbors within a radial cutoff, and a second sum over neighbors defined in SPHC space, which is local in the learned coordinate representation but can couple atoms that are distant in real space.

Other nonlocal architectures where an atom can influence any other atom in a structure are those that leverage global Fourier modes. For example,
Neural P$^3$M~\cite{wang2024neural} combines local geometric graph neural network (GNN) features with a global mesh-based pathway, allowing atomic representations to depend on the entire system configuration through Fourier-domain updates. The model integrates a conventional geometric GNN for the short-range interactions with a separate mesh-based long-range block: atoms (or, better, charges) are assigned to a mesh grid (Atom$\to$Mesh), the mesh representations undergo a Fourier-based long-range update (Mesh$\to$Mesh), and information is back-projected from the mesh to atoms (Mesh$\to$Atom). Because the mesh representation update is computed globally (Fourier space) and its outputs are injected back into the per-atom features, 
the resulting atomic descriptors 
carry genuine nonlocal information.
Another related approach is the Ewald-based Message Passing (Ewald-MP) scheme \cite{kosmala2023ewald}, which augments standard local GNN layers with a reciprocal-space coupling analogous to an Ewald summation. Here atomic features are transformed into Fourier modes, filtered by learnable spectral kernels, and mapped back to real space, thereby enabling long-range correlations to propagate beyond a fixed cutoff. Ewald-MP enhances message passing with a global Fourier-domain pathway, and therefore also produces atom-wise descriptors that depend on the entire configuration.

As a final example of nonlocal ML architecture, an additional recent development is LOREM~\cite{rumiantsev2025learning}, which introduces equivariant charge spherical \textit{tensors} -- rather than scalars -- that participate in a long-range message-passing mechanism. The tensorial character of these equivariant ``charges'' carries higher-order geometric information, and can be used to compute orientation-dependent interactions that go beyond simple distance-based decay laws. Numerical experiments on cumulene, a demanding benchmark for long-range \textit{and} nonlocal effects, show that LOREM does not require large cutoffs or stacking several message-passing layers: the learned long-range messages intrinsically propagate over the entire structure, which allows the model to robustly encode nonlocal effects 
in a way that is both expressive and computationally efficient.

By and large, all the models previously described -- including those based on nonlocal input representations -- can, in principle, be used to learn patterns reminiscent of polarization effects only if they are present in the training dataset. On the one hand, this aspect appears as a natural consequence of avoiding undertaking a self-consistent step at inference; on the other hand, the use of nonlocal representations and/or architectures can be seen as limiting the model transferability when compared to charge-equilibration schemes -- where the nonlocality of the charge distribution emerges as a derived property of the variational procedure. In this context, distinguishing between these two learning paradigms -- and understanding the implications for the model expressivity and transferability -- represents an important direction for future research.

Importantly, the absence of an explicit charge-equilibration mechanism or nonlocal structural information does not imply that all nonlocal electrostatic effects are necessarily inaccessible to purely local models. A question arises whether intrinsically nonlocal physical interactions can be effectively encoded in local representations through correlations present in the training data, even when no nonlocal degrees of freedom are introduced at inference. For example, a molecule with a permanent electric charge (or dipole) facing a metallic slab experiences the interaction with its \textit{image charge} (or image dipole), an induced-polarization effect that is intrinsically nonlocal. In principle, nonlocal effects cannot be captured by models that assign  charges based only on the local environment and without any charge-equilibration mechanism. However, for this geometry, a local  charge model trained on data from that specific system can \textit{effectively} learn the resulting molecule--image-dipole interaction. This learned behavior is system-specific: if the training set  is too diversified, the model will learn only an average interaction across the training dataset that is not accurate for any individual system.
Although this aspect limits the transferability of similar models to capture any sort of long-range polarization phenomenon at inference, it considerably enlarges the spectrum of interactions that can be captured beyond what could be loosely defined as ``permanent'' electrostatics. However, systematic studies investigating the extent to which nonlocal effects can be effectively learned (and transferred) in these contexts are still largely lacking. Related considerations also emerge when examining examples commonly used to motivate charge-equilibration approaches. For instance, scenarios such as the already discussed behavior of Au$_2$ adsorbed on aluminum-doped MgO, are often presented as paradigmatic cases where long-range electrostatics \textit{and} charge redistribution play a decisive role. However, local-charge models trained on global quantum-mechanical targets can possibly reproduce the qualitative trends observed in these systems. This raises important questions regarding which aspects of the underlying physics are truly being captured, which are being effectively mimicked through correlations present in the training data, and which are genuinely required to achieve predictive accuracy across different regimes. These observations highlight the urgent need for coherent and carefully designed benchmarks, in which one explicitly probes what physical mechanisms are necessary for each system and interaction class, and where the limits of local-environment-based charge models, compared to models including nonlocal structural information can be assessed in a controlled and transparent manner.

\section{Finite-field effects~\label{sec:finite-field}}

A related aspect to the prediction of long-range interactions is that of extending ML methodologies to capture the effect of applied electric fields.
The inclusion of a finite electric field in atomistic simulations becomes essential whenever one aims to describe phenomena that cannot be captured through linear‐response theory or other perturbative treatments. While many dielectric and polarization properties, as well as transport coefficients, can be obtained from zero-field simulations -- either by evaluating response functions or using fluctuation–dissipation/Green-Kubo relations -- important situations exist where the explicit presence of an electric field is indispensable. Examples include field-driven structural transitions~\cite{wang2020electric,chen2025electric}, finite-field stabilization of metastable states~\cite{stocco2025electric}, field-activated catalytic reactions~\cite{Che2018}, molecular dissociations~\cite{Saitta2012,Fidanyan2023} and proton transport~\cite{cassone2014proton}, as well as when the field is used to simulate an applied potential difference at electrochemical interfaces~\cite{futera2021water,Andersson2025}. 

Within ML-driven atomistic simulations, incorporating finite-field effects is particularly challenging because the field couples directly to the electronic structure, whereas standard ML interatomic potentials are trained only on geometry-dependent energies and forces. Two conceptually distinct ML strategies have been developed to overcome this challenge. In the first, one decouples the problem of field-response from the zero-field potential. Specifically, the latter is augmented with separately learned response quantities, such as the polarization, Born effective charges, or polarizabilities, that allow the system to couple to an external field. In the second, a single ML model is trained to represent the field-dependent energy as a joint function of the atomic coordinates and the applied field, thereby capturing finite-field effects explicitly.
In this section, we provide the physical context and comment on fundamental principles underlying the inclusions of finite-field effects in atomistic simulations. 
We then discuss how these ideas are implemented in ML models, by comparing the capabilities and limitations of the two ML strategies previously outlined.

\subsection{Physical context}

When a homogeneous electric field $\boldsymbol{E}$ is applied to a system, its total energy is modified from its zero-field value $U^0$ through the coupling with the system's polarization $\boldsymbol{P}$:
\begin{equation}
    U(\{\boldsymbol{R}_i\},\boldsymbol{E}) = U^0(\{\boldsymbol{R}_i\}) - \Omega \, \boldsymbol{P}(\{\boldsymbol{R}_i\},\boldsymbol{E}) \cdot \boldsymbol{E}\, ,
\end{equation}
sometimes referred to as the \textit{electric enthalpy}. The corresponding field-dependent forces are given by
\begin{equation}
    \boldsymbol{F}_i = -\frac{\partial U}{\partial \boldsymbol{R}_i}  = \boldsymbol{F}_i^0 + \mathbf{Z}_i \cdot \boldsymbol{E}, \label{eq:finE_forces}
\end{equation}
where $\mathbf{Z}_i$ denotes the Born effective charge (BEC) tensor of atom $i$.  
These relations imply the use of the following definitions for their Cartesian components:
\begin{equation}\label{eq:ff-def}
    \Omega\, P_\alpha = -\frac{\partial U}{\partial E_\alpha}, \qquad 
    Z_{i,\alpha\beta} = -\frac{\partial^2 U}{\partial E_{\alpha}\, \partial R_{i,\beta}}, 
\end{equation}
In practice, a linear-response approximation is typically adopted when working in the regime of relatively small applied fields, so that energies and atomic forces are expressed from the polarization and BECs at zero field, respectively:
\begin{equation}
 U \approx U^0 - \Omega \, \boldsymbol{P}^0 \cdot \boldsymbol{E}\, ,\qquad   \boldsymbol{F}_i \approx \boldsymbol{F}_i^0 + \mathbf{Z}_i^0 \cdot \boldsymbol{E}\, . \label{eq:finE_forces_E=0}
\end{equation}
At large field intensities, second-order approximations of the electronic energy can further be derived by approximating the polarization vector at first order from the the electronic polarizability tensor $\boldsymbol{\upalpha} = \partial \boldsymbol{P}/\partial{\boldsymbol{E}}$, 
\begin{equation}
    \boldsymbol{P}(\boldsymbol{E}) \approx \boldsymbol{P}^0 + \boldsymbol{\upalpha}^0 \cdot \boldsymbol{E},
\end{equation}
implying that the corresponding contribution to the forces is described by a third-order atomic tensor.

\subsection{Conservation laws and consistency conditions}

The definitions of Eq.~\eqref{eq:ff-def} impose a set of important conservation laws.  
First, since the polarization is the derivative of the energy with respect to the electric field, the finite-field work computed over a closed path must vanish:
\begin{equation}
    \oint \boldsymbol{P} \cdot \mathrm{d}\boldsymbol{E} = 0. \label{eq:line_int_P_conservation}
\end{equation} 
Similarly, since the BECs are given by the derivative of the polarization with respect to atomic displacements, the transported (adiabatic) charge associated to a path $\gamma$, starting from any point in the space of atomic configurations and ending to an image of the original point in a replica of the simulation cell identified by $(l a_1, m a_2, n a_3)$, where $a_1$, $a_2$, and $a_3$ are the lengths of the three lattice vector of the simulation cell, here assumed orthorhombic for simplicity, is given by:
\begin{equation}
    \int_\gamma \sum_{i=1}^N \mathbf{Z}_i \cdot \mathrm{d}\boldsymbol{R} 
    = \Omega\int_\gamma \mathrm{d}\boldsymbol{P} = (Q_1 a_1, Q_2 a_2, Q_3 a_3)\, ,\label{eq:integer_int_dP}
\end{equation}
with $Q_1$, $Q_2$, and $Q_3$ that are \textit{integer numbers}.
This relation has been used to provide topological definitions of the atomic oxidation numbers, where their integer-ness naturally emerges as a fundamental physical property~\cite{JiangLevchenkoRappe2012,grasselli2019topological,pegolo2020oxidation}. Furthermore, Eqs.~\eqref{eq:line_int_P_conservation} and \eqref{eq:integer_int_dP} say that under the effect of an adiabatically varying electric field, the hysteresis loop closes exactly, meaning that the path traced by the polarization in the $\boldsymbol{P}$–$\boldsymbol{E}$ plane forms a closed curve: once the electric field is cycled back to its initial value, the polarization also returns to its initial value without any residual offset, as explicitly observed for BaTiO$_3$ in Ref.~\cite{Falletta2025}.
Finally, the invariance of the polarization under a constant shift $\boldsymbol{u}$ of all nuclear positions (which follows from the translational invariance of the electronic energy) implies the \textit{acoustic sum rule} \cite{grosso2013solid}:
\begin{equation}
    \Omega\,  \frac{\partial \boldsymbol{P}}{\partial \boldsymbol{u}} = \sum_{i=1}^N \mathbf{Z}_i = \boldsymbol{0}. \label{eq:acoustic_sum_rule}
\end{equation}
Importantly, when $\boldsymbol{P}$ or $\mathbf{Z}_i$ are predicted by a ML model, rather than computed as derivatives of an energy function, these conservation laws may not be automatically satisfied, and they must therefore be explicitly enforced.

\subsection{Decoupled (direct-response) models}

\begin{figure}[h!]
    \centering
    \includegraphics[width=0.7\linewidth]{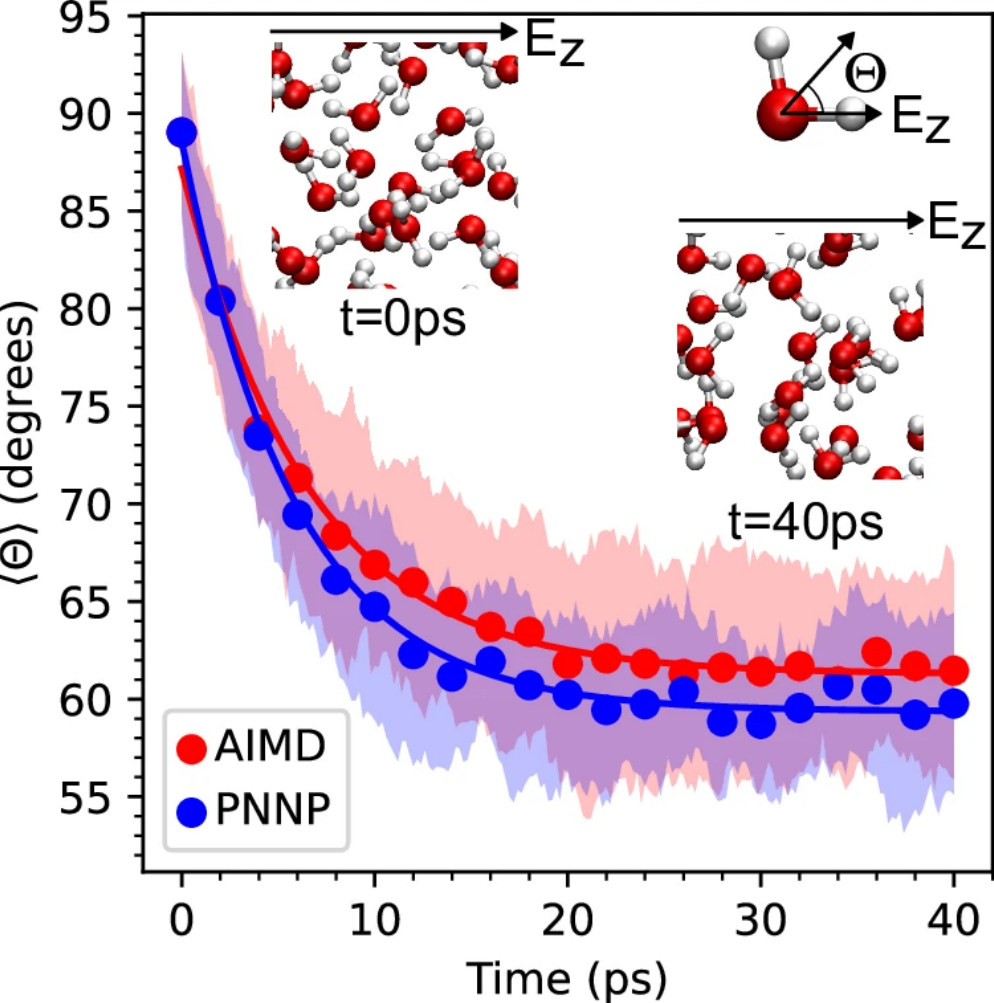}
    \caption{{The change in the average orientation of the water molecules defined by $\langle \Theta \rangle$ is shown for the Perturbed Neural Network Potential (PNNP) molecular dynamics (blue circles), and \textit{ab initio} molecular dynamics (AIMD) (red circles) after an electric field along the $z$-direction with field strength $E_z = 0.0257~\mathrm{V\,\AA^{-1}}$ is switched on at $t = 0$. The angle $\Theta$ is defined for each water molecule as the angle between the bisector of the two intramolecular OH bond vectors and the applied electric field vector, as illustrated in the inset at the top right corner. Reproduced from Ref.~\cite{Joll2024}, licensed under CC~BY~4.0.}}
    \label{fig:pnnp}
\end{figure}

In decoupled (direct-response) approaches, the zero-field  energy  $U^0$ and its corresponding forces are learned independently of the field-response properties. %
In particular, tensorial models of the BECs required in Eq.~\eqref{eq:finE_forces_E=0} are separately trained using reference \textit{ab initio} data obtained either from finite differences of finite-field calculations~\cite{umari+02prl}, or from density-functional perturbation theory~\cite{baroni2001}.  
This approach has been adopted, for instance, in Ref.~\cite{Joll2024}, which uses E(3)-equivariant graph neural networks to learn the $\mathbf{Z}^0_i$, and then applies the method to compute several dielectric properties of liquid water under a finite field {(Fig.~\ref{fig:pnnp})}.  
Because the response tensors are not learned as derivatives of an underlying potential, the conservation laws  previously outlined do not automatically hold: the acoustic sum rule of Eq.~\eqref{eq:acoustic_sum_rule} must be explicitly imposed, and the field-dependent force field is not guaranteed to remain conservative. A strategy to restore conservativity is to complement the zero-field potential with a separate model for the polarization vector.  
For example, Ref.~\cite{zhan+20prb} combines a conventional ML potential with a $\boldsymbol{P}$ value obtained from predicted Wannier centers.  
Although such a formulation is, in principle, compatible with the inclusion of finite electric fields, this capability has not yet been demonstrated in simulations.  
Conversely, in Ref.~\cite{shimizu2023prediction} the macroscopic polarization is directly regressed through a differentiable model, from which BECs follow as exact derivatives,
thereby ensuring conservative forces even under finite applied fields.
This latter method, in particular, has been applied to simulate the mean square displacement of Li$^+$ ions in crystalline and amorphous solid-state electrolyte Li$_3$PO$_4$ under a finite field.

Beside explicit finite-field simulations, a broad class of studies instead infer dielectric polarization responses from equilibrium simulations via linear-response formalism and leveraging correlation functions, thereby avoiding the need to impose external fields directly in simulations \cite{all-tild90book,frenk-smit02book}. In this framework, the relevant response functions emerge from the time correlations of spontaneous fluctuations \cite{kubo66rpp}, which can be sampled from equilibrium simulations and fully characterize the linear response.

\begin{figure}
    \centering
    \includegraphics[width=0.8\linewidth]{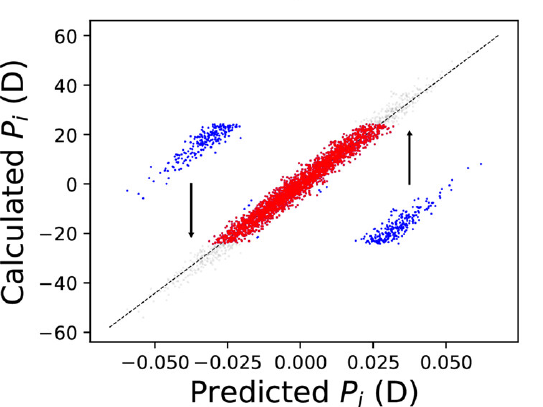}
    \caption{{Blue crosses show the total polarizations predicted using a model trained on their derivatives. Points in red are those identified as the ``main branch'', which is fit to the straight line shown. All points are then shifted by $\boldsymbol{Q}\boldsymbol{n}$ to be as close as possible to the straight line, where $\boldsymbol{Q}$ is the quantum of polarization and $\boldsymbol{n}$ is a vector of integers; this is illustrated for points off of the main branch by arrows; their shifted positions are shown in gray. Reproduced from Ref.~\cite{jana2024learning}, licensed under CC~BY~4.0.}}
    \label{fig:branches}
\end{figure}

In practice, predicting the polarization of periodic systems with ML models requires careful handling of the multivalued nature of the Berry-phase polarization.  
A popular and flexible strategy is to represent the macroscopic polarization as a sum of local, environment-dependent dipoles,
\begin{equation}
    \boldsymbol{P}=\frac{1}{\Omega}\sum_{i=1}^N\boldsymbol{d}_i\, ,
\label{eq:polarization_sum_local_contributions_merged}
\end{equation}
hence exploiting electronic nearsightedness principles in analogy with the local decompositions of electronic energies.  
To render Eq.~\eqref{eq:polarization_sum_local_contributions_merged} a well-posed regression problem, one must first select a single branch of the Berry-phase polarization, folding all raw DFT values onto a consistent main branch.  
This procedure has been streamlined in
Ref.~\cite{jana2024learning}, where an efficient preprocessing method based on a small set of polarization derivatives is implemented.  
Such a strategy then makes it possible to train directly on $\boldsymbol{P}$ while avoiding multivaluedness {(Fig.~\ref{fig:branches})}. The problem of the multivalued nature of the polarization was also recently addressed in Ref.~\cite{stocco2025electric} by ensuring that branch jumps inherent to periodic systems are treated consistently by the ML model, enabling the simulation of phase transitions in a ferroelectric perovskite, together with its excitation under THz electric fields of varying frequency and intensity.

An alternative possibility, particularly useful in ferroelectric materials, is to target the ``effective'' polarization~\cite{resta2007theory}, defined as the difference with respect to a centrosymmetric reference configuration, thus removing branch ambiguities.  
If the ions are displaced by $\Delta \boldsymbol{R}$ from the reference, the effective \textit{ab initio} polarization to be used as target can be written as
\begin{equation}
    \Delta \boldsymbol{P}(\Delta \boldsymbol{R})
    = \frac{1}{\Delta \boldsymbol{R}}\int_0^{\Delta \boldsymbol{R}} \sum_{i=1}^N
    \mathbf{Z}_i(\boldsymbol{R})\cdot \mathrm{d}\boldsymbol{R}\, .
\end{equation}
This yields a continuous single-valued target suitable for regression via Eq.~\eqref{eq:polarization_sum_local_contributions_merged}, as it is used, for instance, for predicting the  polarization fields governing the ferroelectric transitions in  BaTiO$_3$~\cite{gigli2022thermodynamics}.

\subsection{Unified energy-based models}

Unlike the models just discussed, an alternative strategy to include finite-field effects is to learn a single e model for $U(\boldsymbol{R},\boldsymbol{E})$, from which all response properties can be consistently derived.
In this framework, the physical consistency between polarization, BECs, and polarizabilities is ensured by construction. Ref.~\cite{ZhangJiang2023} follows this approach, training $U(\boldsymbol{R},\boldsymbol{E})$ on \textit{ab initio} data computed under finite electric fields.  
Once again, since the polarization of a periodic system is only defined modulo a quantum of polarization, the targets were ``folded'' onto a single branch to eliminate branch discontinuities that cannot be handled by single-valued functional models as neural networks. The approach is implemented in the field-induced recursively embedded atom neural network (FIREANN) model, and tested on field-free and in-field vibrational spectra of molecular N-methylacetamide and liquid water.

\begin{figure}[h!]
    \centering
    \includegraphics[width=0.8\linewidth]{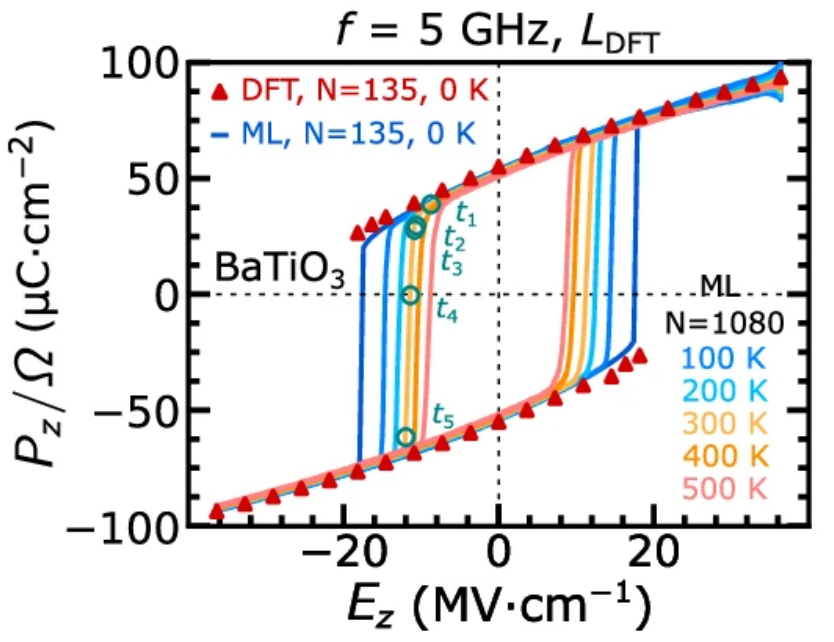}
    \caption{{Ferroelectric hysteresis of polarization density $P_z/\Omega$ as a function of an applied electric field along $z$ in BaTiO$_3$, as obtained using DFT-optimized lattice parameters (LDFT). This includes results for a 135-atom supercell at $0~\mathrm{K}$, obtained through structural relaxations with ML or DFT, and results for a 1080-atom supercell at finite temperatures, calculated through ML molecular dynamics. Adapted from Ref.~\cite{Falletta2025}, licensed under CC BY-NC-ND 4.0.}}
    \label{fig:histeresis}
\end{figure}

Interestingly, Ref.~\cite{Falletta2025} solves the problem of preprocessing the unfolded DFT polarization by leveraging modular algebra.  
In particular, a loss term related to polarization is here introduced as,
\begin{equation}
    \mathcal{L}_{\boldsymbol{P}}({\boldsymbol{\theta}}) \propto \sum_{\alpha=1}^3
    \left|
        \left(
            -\frac{\partial U({\boldsymbol{\theta}})}{\partial E_\alpha} 
            - P^{\mathrm{ref}}_\alpha
        \right)
        \mathrm{mod} \frac{a_\alpha}{\Omega}
    \right|,
\end{equation}
where the “$\mathrm{mod}$” operation involves a minimum-image convention in polarization space with respect to the polarization quanta defined from the unit cell translation vectors of length $a_1$, $a_2$ and $a_3$.  
Further terms in the loss-function are then added to enforce higher-order consistency:
\begin{align}
    \mathcal{L}_{\mathbf{Z}}({\boldsymbol{\theta}}) &\propto 
        \sum_{i=1}^N \sum_{\alpha=1}^3\sum_{\beta=1}^3\left|-\frac{\partial^2 U({\boldsymbol{\theta}})}{\partial E_\alpha\, \partial R_{i,\alpha\beta}} 
        - Z_{i,\alpha\beta}^{\mathrm{ref}}\right|, \\
    \mathcal{L}_{\boldsymbol{\upalpha}}({\boldsymbol{\theta}}) &\propto \sum_{\alpha=1}^3\sum_{\beta=1}^3
        \left|-\frac{\partial^2 U({\boldsymbol{\theta}})}{\partial E_\alpha \partial E_\beta} 
        - \upalpha_{\alpha\beta}^{\mathrm{ref}}\right|.
\end{align}
Here, finite-field data are used only to compute derivatives (e.g., BECs and polarizabilities) at zero field, typically via finite differences. 
This framework has been applied to $\alpha$-SiO$_2$ to predict vibrational spectra and dielectric response, and to the ferroelectric material BaTiO$_3$ to reproduce the evolution of polarization under varying temperature and electric-field conditions {(Fig.~\ref{fig:histeresis})}.

We note that most of the aforementioned ML approaches rely on  local atomic-environment representations, and therefore do not include long-range physical interactions. If one is only interested in instantaneous electronic responses, this assumption can be justified by the negligibly small value of the applied field that is normally introduced to impose an electric bias in the system. However, whenever the field builds up an induced macroscopic polarization during the molecular dynamics simulation, either as a consequence of specific charge migrations or dipolar  alignments, the system will necessarily acquire a marked long-range physical character that cannot generally be disregarded in the ML approximation of the electronic energy. In this context, a few long-range ML models have already been proposed that explicitly allow for finite-field coupling. These include self-consistent models that explicitly include the external field as an input parameter of the learning architecture~\cite{Gao2022}, implicit polarization models of the long-range energy that include computed values of BECs in the loss-function~\cite{li-scandolo2026}, {approaches where polarization and BECs are extracted from the implicit charges learned from the long-range contribution trained solely on energies and forces
~\cite{zhong2025machine},} as well as ML charge-equilibration schemes where the effect of applied fields is introduced as an additional term of the QEq energy~\cite{Vondrak2025}.

\section{Discussion: electrochemical interfaces and ionic transport~\label{sec:discussion}}

The discussion carried out during the previous sections rationalizes different learning paradigms of electrostatic interactions and finite-field effects following fundamental physical principles. We believe that a similar line of reasoning will be important for guiding the inclusion of long-range phenomena within ML-driven simulations, especially in those contexts governed by a nontrivial interplay of electrostatics, polarization, and charge-transfer processes.  
A prototypical example is given by electrochemical interfaces. Here, the mutual influence between the electrode’s electronic structure, the solvation environment, and the distribution of ionic species gives rise to the so-called \textit{electrical double layer} (EDL), representing the physical background where energy storage in batteries and supercapacitors take place. Capturing such effects requires models capable of resolving collective fluctuations, field-induced responses, and dynamic charge redistribution, dictating the thermodynamic and kinetic properties of the system across multiple length and time scales~\cite{Zhang2020,Jinnouchi2025}.

\begin{figure}[h!]
    \centering
    \includegraphics[width=0.9\linewidth]{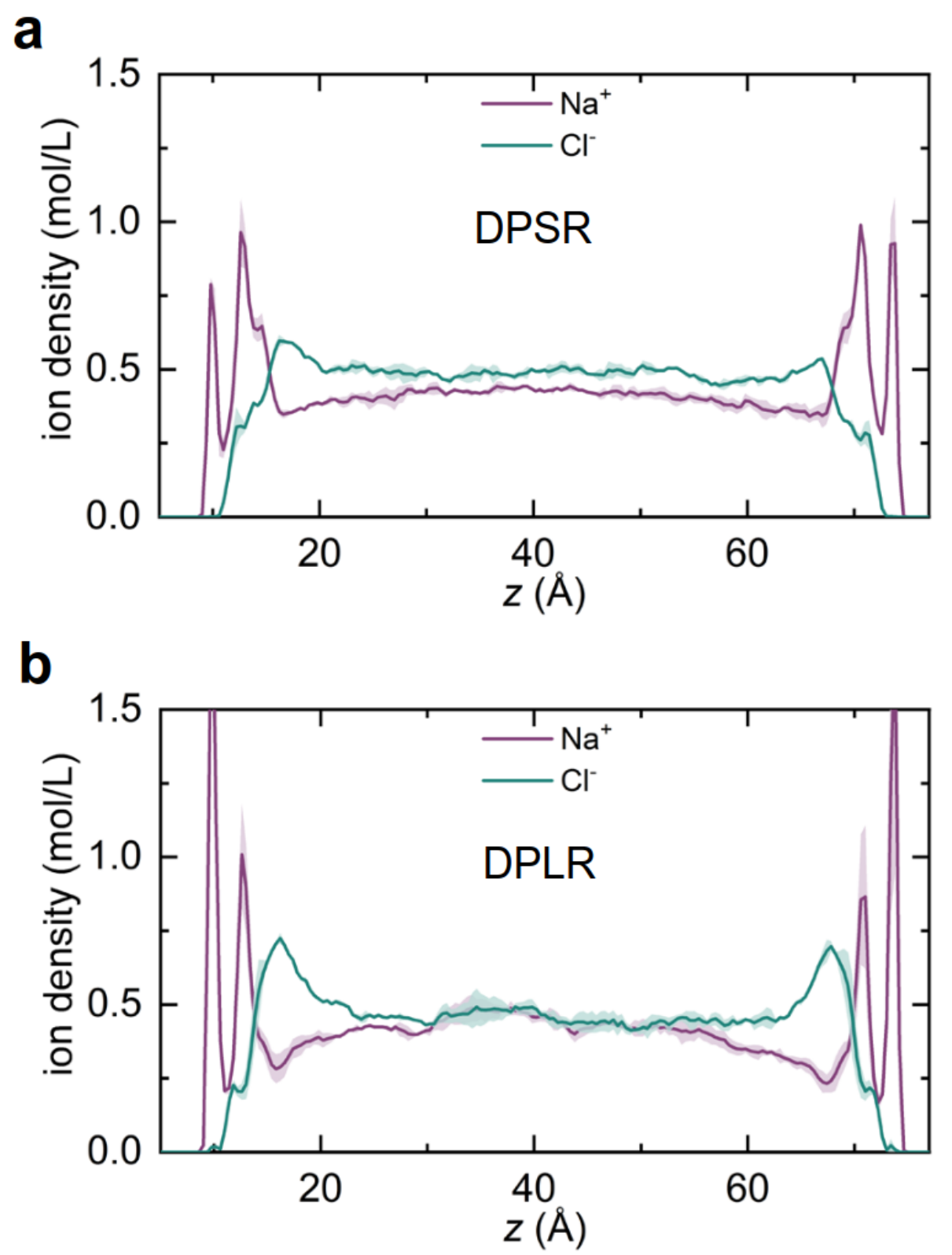}
    \caption{{Plane-averaged ion distributions along the $z$-direction for the TiO$_2$/NaCl(aq) interface obtained from short-range (a) and long-range (b) ML potentials over 5~ns simulations. Adapted from Ref.~\cite{Zhang2024natcomm}, licensed under CC BY-NC-ND 4.0.}}
    \label{fig:screening}
\end{figure}

A key aspect towards achieving realistic simulations of the EDL structure is that of reproducing the perfect electrostatic screening of the excess charge accumulated at the electrode surface, responsible for establishing a vanishing electric field in the electrolyte bulk region. This phenomenon originates from the distribution of ionic charge carriers at thermodynamic equilibrium and it typically requires collecting statistics over long simulation trajectories of several  nanoseconds~\cite{dufils+19prl}. In turn, this allows one to retrieve the expected Debye-like exponential decay of the charge density towards the liquid bulk, which, depending on the temperature and ionic concentration, can cover several nanometers from the electrode surface. Unlike in the simulation of bulk ionic liquids and solutions, the inherent statistical anisotropy of the electrostatic fields in the system -- derived by the presence of electrode/electrolyte interface -- implies that this screening effect \textit{cannot} be captured without an explicit inclusion of long-range interactions in ML potentials. This has been shown, for instance, in the context of simulating the electrified TiO$_2$/electrolyte interface at varying H$^+$ coverage~\cite{Zhang2024natcomm}, where it is proven how a short-range ML model alone is not capable of achieving ionic charge-neutrality in the electrolyte bulk~{(Fig.~\ref{fig:screening})}. In this case, perfect ionic screening is recovered upon inclusion of long-range interactions via Wannier-centers representations discussed in Sec.~\ref{sec:wannier}. Interestingly, the same result has also been later reproduced by the LES model discussed in Sec.~\ref{sec:implicit}, remarking how this phenomenon is fundamentally driven by ionic charge migrations and can, in general, be well captured by  local ML approximations of the charges in the system. A similar argument also applies to the orientational screening of surface dipoles that are spontaneously formed at an atomistic interface involving a dielectric medium like water, as proven multiple times in the context of liquid-vapor interfaces~\cite{Niblett2021,Cheng2025,li-scandolo2026}.

\begin{figure}[h!]
    \centering
    \includegraphics[width=1\linewidth]{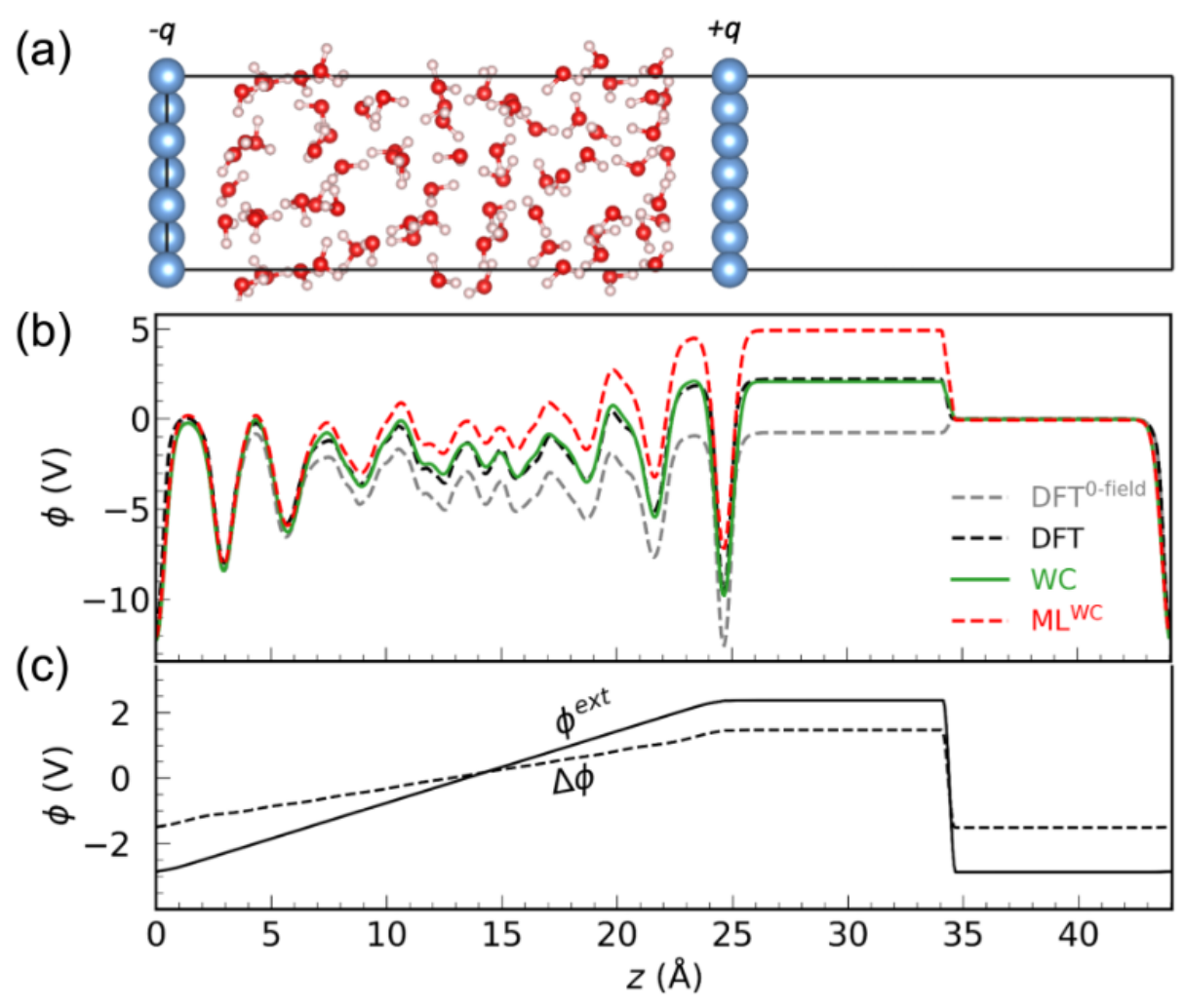}
    \caption{{(a) Atomistic model of the nanoconfined water between
two oppositely charged Ne electrodes. (b) The
corresponding electrostatic potential calculated from DFT, WCs charges, and a long-range ML model based on WCs (ML$^\text{WC}$), with an applied field of $E_\text{ext} = 0.2$~V/{\AA}. The gray
dashed line represents the zero field DFT reference. (c) The
externally applied potential $\phi_\text{ext}$ and the difference between the DFT
electrostatic potential with and without field $\Delta\phi$. In the bulk
water region, $\Delta\phi$ corresponds to an electric field smaller than
$E_\text{ext}$, which results from the electronic screening of water. Reproduced from Ref.~\cite{Yang2026prl}, licensed under CC~BY~4.0.}}
    \label{fig:dielectric}
\end{figure}

Beyond ionic and thermal dipolar screening, an important (still largely disregarded) aspect in applications of long-range ML models to solid/electrolyte interfaces is that of reproducing the \textit{macroscopic}  electronic polarization that naturally emerges in instantaneous  configurations of the    electronically insulating portion of the system. This phenomenon has been recently characterized in Ref.~\cite{Yang2026prl}, by considering a slab of liquid water between two (possibly charged) walls as a minimal model of an open-circuit capacitor. It was observed how the component of the total polarization vector along the direction orthogonal to surface, $P^\text{tot}_z$, is strongly affected by the collective electronic dielectric screening built up from the polarization of the individual water molecules. In turn, this effect was shown to carry a noticeable impact on the $z$-variations of electrostatic potential across the electrochemical interface, representing a crucial physical quantity for the calculation of work functions and electrode potentials. {As shown in Fig.~\ref{fig:dielectric},} in particular, a ML partitioning of the charges in the system based on local predictions of Wannier centers was not capable of reproducing the underlying global electronic response, yielding an error in $P^\text{tot}_z$ commensurate with the expected value of electronic dielectric constant in liquid water, i.e., $\varepsilon^\text{w}_\infty \approx 1.78$. This example is especially striking when considering that the Wannier-centers construction derives from a consistent definition of the polarization in periodic system (Sec.~\ref{sec:wannier}). Yet, this property alone is not sufficient for guaranteeing that the model can capture nonlocal polarization effects, for which predicting charge distributions based solely on local structural information are not appropriate.

\begin{figure}
    \centering
    \includegraphics[width=0.8\linewidth]{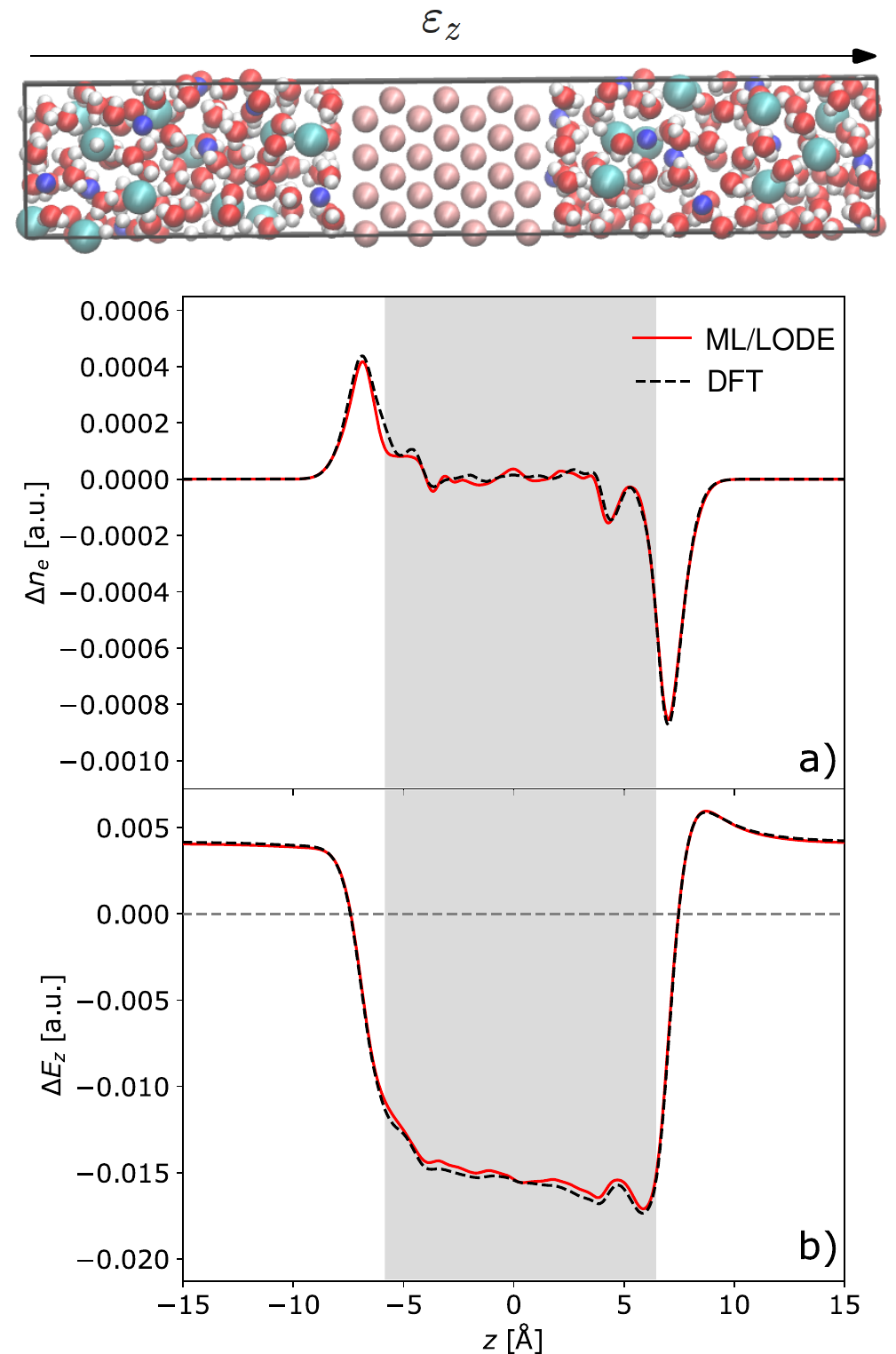}
    \caption{{Top: representation of an ionic capacitor made of a Au(100) electrode put in contact with a NaCl solution under 3D periodic boundary conditions. According to Ref.~\citenum{Dufils2019}, a finite electric field $\varepsilon_z$ is applied along $z$ to mimic a voltage drop of $\Delta V=-\varepsilon_z L_z$ between two parallel electrodes. Bottom: electrode charge density response (a) and derived electric field (b) induced by the classical charges of a test electrolyte configuration. Red lines: ML/LODE prediction. Black dashed lines: DFT reference. Shaded gray area indicates the region occupied by the gold electrode. Adapted from Ref.~\cite{Grisafi2024}. Copyright 2024 AIP Publishing.}}
    \label{fig:electrode}
\end{figure}

Of course, a complementary aspect where nonlocal polarizations are more transparently manifested is the description of the electrode's electronic response. A consolidated strategy to account for this effect in classical electrochemical simulations is that of adopting QEq schemes that are specifically formulated to enforce a constant-potential bias in the system~\cite{Willard2009,Limmer2013,Dufils2023} -- historically referred to as the Siepmann-Sprik model~\cite{Siepmann1995}. Indeed, the electronically conductive character of typical electrode materials implies that the CPE principle discussed in Sec.~\ref{sec:QEq} is physically grounded, thus making classical QEq models generally suitable for describing nonreactive electrochemical interfaces such as those encountered in ionic capacitors~\cite{Jeanmairet2022}. For this reason, hybrid descriptions of electrochemical interfaces have recently been proposed where the Siepmann-Sprik electrode model is integrated with conventional ML potentials for the short-range interactions, and further coupled with a long-range electrostatic term derived from local ML predictions of the electrolyte charges~\cite{Zhu2025,wang2025cheng}. Importantly, the fact that the short-range part of the ML potential also includes the electrode/electrolyte interactions here compensates for the simplicity of the Siepmann-Sprik model, effectively absorbing the missing physics in the description of the electrode surface charge. In particular, it is well know how classical electrode models lack a description of \textit{electronic spillover}, manifested as a diffuse spatial response of the electronic charge-density that goes from right outside the metal surface deep into the electrolyte phase~\cite{Schmickler1996,Li2025jacs}. Dedicated classical corrections to the Siepmann-Sprik model have in fact been recently proposed that account for this effect through the inclusion of surface dipoles terms, demonstrating the capability of the model of recovering the experimental capacitance-voltage curve of a prototypical metal/electrolyte interface~\cite{Wang2025}. In this context, a complementary ML approach to those previously discussed is that constructing an electron-density model of the electrode charge that can naturally incorporate the spillover physics included in DFT calculations. This strategy has been undertaken in Ref.~\cite{Grisafi2024} by embedding a quantum-based description of the electrode within the classical environment of a concentrated electrolyte solution. Crucially, LODE input features as discussed in Sec.~\ref{sec:nonlocal-rep} are here used by the ML model to predict the nonlocal electronic polarization of the electrode charge density as a function of the electrolyte atomic positions, achieving remarkable accuracy in the derived calculation of electrostatic forces {(Fig.~\ref{fig:electrode}).} Yet, extensions of the method that also consider part of the electrolyte solution as participating in the interfacial response will be required to fully account for the electronic redistributions deep into the liquid phase. 

A specific mention is due to the problem of simulating a constant potential difference via the coupling with an external field. In fact, while a popular strategy in first-principles electrochemical simulations makes use of grand-canonical DFT methods to fix the electronic chemical potential in the electrode bulk~\cite{Sundararaman2017,Melander2019,Hoermann2019}, the net fluctuations of the metal surface charge necessarily imply adopting continuum descriptions of the electrolyte solution capable of enforcing global electroneutrality of the simulation cell. In this context, 
application of an external electric field $E_z$ along the direction orthogonal to the electrochemical interface, such that $\Delta V = - E_z L_z$ corresponds to a physical cell potential, can alternatively be used to preserve an explicit charge-neutral description of the electrolyte solution~\cite{umari+02prl,Stengel2007,Zhang2016prb-2,Dufils2019}. Similarly, an open-circuit condition representing a constant electrode surface-charge can be simulated through the application of a finite electric displacement~\cite{Stengel20009,Zhang2020,Andersson2025}. Refs.~\cite{Zhu2025,wang2025cheng,Grisafi2024} previously mentioned are all examples of ML-driven simulations of electrochemical interfaces that follow this rationale by explicitly introducing a uniform electric field through the system via the coupling with the system's polarization. However, different strategies that adopt the FIREANN method reported in Sec.~\ref{sec:finite-field} have recently been presented where the external field becomes an integral part of the descriptor, thus enabling learning the full finite-field potential energy surface of the electrochemical interface directly, by also including predictions of the electrode charge-density response~\cite{feng2025arxiv}. Besides finite-field simulations, constant-charge perturbations have also been recently explored to machine learn the energetics of metal-liquid interfaces. Reference \cite{Bergmann2025} introduce a local ML potential that learns how the energy and forces of an electrified interface change as a function of a finite imposed charge on a confined slab. Hence, instead of treating the electric field as the external input variable as in Refs. \cite{ZhangJiang2023,Falletta2025}, the perturbation parameter is the excess bias charge $q$. This let the model  capture work-function shifts and charge-dependent force responses at electrodes. Conceptually, this extends the same strategy of embedding response physics directly into a local ML potential, but applied to constant-charge rather than finite-field conditions.

Beyond electrostatic properties, an additional relevant question in the simulation of electrochemical systems is that of understanding the role of long-range interactions in the calculation of transport properties, and in particular of ionic conductivities to determine the charging performance of energy devices. 
Charge-density fluctuations are, in general, highly sensitive to the accurate treatment of long-range electrostatics, since their behavior in ionic systems is governed by perfect Coulomb screening at long wavelengths. In striking contrast, most practical calculations of ionic electrical conductivity rely on the Green-Kubo (GK) integral of the autocorrelation function of the charge flux, 
\begin{equation} 
\sigma = \frac{\Omega}{3k_\mathrm{B}T}\int_0^\infty \langle \boldsymbol{J}(t)\cdot\boldsymbol{J}(0)\rangle\,\mathrm{d}t. \label{eq:sigma_from_GK} \end{equation}
For this approach it has been shown that the resulting conductivity is in practice only weakly affected by whether long-range forces are included explicitly or approximated by a short-range model (see Ref.~\cite{Staacke2022}). The literature on this point is still limited, and the physical reason behind the discrepancy between the sensitivity of density versus current fluctuations remains under investigation. %
The microscopic definition of charge flux requires particular care. In first-principles or polarizable models, the charge flux is rigorously defined using the Born effective charge tensors of the ions:
\begin{equation} 
\boldsymbol{J}(t) = \frac{1}{\Omega}\sum_{i=1}^N \mathbf{Z}_i(t) \cdot \dot{\boldsymbol{R}}_i(t) \label{eq:J_quant}
\end{equation}
However, thanks to the gauge-invariance principle for transport coefficients and some arguments from quantization of particle transport \cite{grasselli2019topological,pegolo2020oxidation}, under rather general assumptions (such as the absence of dissolved electrons or polarons) one can replace these Born charges with the ions' time-independent, integer oxidation numbers $z_i \in \mathbb{Z}$. For instance, for a stoichiometric NaCl molten salt, $z_\text{Na}=+1$ and $z_\text{Cl}=-1$. 
This amounts to obtaining a different microscopic flux:
\begin{equation} \boldsymbol{J}'(t) = \frac{1}{\Omega}\sum_{i=1}^N z_i \, \dot{\boldsymbol{R}}_i(t)\, , 
\end{equation}
but \textit{exactly} the same electrical conductivity as that given by using the flux $\boldsymbol{J}(t)$ defined in Eq.~\eqref{eq:J_quant}. This greatly simplifies GK conductivity calculations.
It is important, however, not to generalize this principle to arbitrary ``atomic charges'', such as those produced by machine-learning interatomic potentials. Such charges do not possess the physical properties of Born charges nor those of oxidation numbers, and there is no guarantee -- nor theoretical basis -- to assume that inserting them into the flux definition will yield the correct conductivity. In ML potentials, conductivity should therefore be computed using oxidation numbers (when applicable) or alternative well-defined gauge-invariant constructions, rather than relying on model-specific atomic charges.

\section*{Acknowledgments}
The Authors are grateful to Michele Ceriotti, Enrico Drigo, Marco Gibertini, Marcel Langer, Philip Loche, and Benjamin Rotenberg for providing useful comments on an early version of the manuscript.
This article is based upon work from COST Action CA22154 - Data-driven Applications towards the Engineering of functional Materials: an Open Network (DAEMON) supported by COST
(European Cooperation in Science and Technology).

%apsrmp4-2.bst 2018-12-27 (MD) hand-edited version of apsrmp4-1.bst
%Control: author (3) reversed first dotless

\end{document}